\documentclass[12pt]{JHEP3}
\usepackage{amsfonts,amssymb}
\usepackage{amsmath}

\usepackage{graphicx}

\newcommand{\BC}{{\mathbb{C}}}
\newcommand{\BZ}{{\mathbb{Z}}}

\newcommand{\tr}{\hbox{tr}}

\title{Monopole operators, moduli spaces and dualities. }
\author{David Berenstein$^{\dagger\ddagger}$, Mauricio Romo$^\dagger$ \\
$\!^\dagger$ Department of Physics, UCSB, Santa Barbara, CA 93106\\
$\!^\ddagger$ School of Natural Sciences, Institute for Advanced Study, Princeton, NJ 08540}

\abstract{We develop semiclassical methods to analyze the spectrum of BPS monopole operators for superconformal field theories in three dimensions with $\mathcal{N}=2$ supersymmetry. We show that the chiral ring of the theory results from the semiclassical holomorphic quantization of the solution of classical BPS equations of motion on the cylinder. We apply this formalism to various theories. We also use these techniques to compare the chiral rings of theories that might be related to each other via Seiberg dualities in four dimensions in the classical limit. We find that the change of basis transformations that
generate dualities in four dimensions (homological operations) generically do not work in three dimensions in the presence of Chern-Simons terms. Instead, new theories
generally arise this way. When dualities are possible, the Chern-Simons couplings need to satisfy certain arithmetic congruences. We also determine the spectrum of R-charges of the chiral ring operators by assembling them on a Hilbert series and by minimizing the coefficient of the maximum pole relative to the trial R-charge. This is related to volume minimization in theories with dual supergravity setups.
}


\begin{document}

\section{Introduction}

Supersymmetric theories in four dimensions can have continuous families of vacua. These families are usually described locally by a complex manifold (the moduli space of vacua). More precisely, they are described by a complex variety. The variety can be reduced or not, and it can have many components. The order parameters that classify the different points in the moduli space of vacua are usually described by operators in the chiral  ring of the theory . Indeed, in all known examples the chiral ring suffices to classify the vacua of these theories. This is encapsulated in the concept of holomorphy which can be used very elegantly to solve for the vacuum structure of  supersymmetric theories (see \cite{Seiberg} for a concise introduction to the subject).
Many field theories that describe different UV physics can have the same moduli space of vacua and thus share a lot of their low energy dynamics. Field theories  that are identified via this low energy universal behavior of the moduli space of vacua are usually called Seiberg dualities \cite{Seibergdual}. Apart from the moduli space of vacua, the theories in four dimensions also have anomalies for global symmetries, and Seiberg dualities require also anomaly matching between the theories \cite{Seibergdual}. This is really important at symmetry points in the moduli space, where the symmetries are unbroken and can be used most effectively.

Furthermore, if one has a superconformal field theory, one has more control over the dynamics. For example, one can compute the R-charge of the theory by the procedure of $a$-maximization \cite{InW}, which is closely related to the anomaly structure of the global symmetries.
In quiver theories one can also have AdS/CFT dual pairs, and the process of a-maximization can be understood in terms of Einstein's equations in the dual theory by the procedure of volume minimization \cite{MSY,Martelli:2006yb}, where one minimizes the volume of a Sasaki-Einstein manifold satisfying some constraints. These two procedures are closely related as proved in \cite{Butti:2005ps,Lee:2006ru,Eager}.

When we consider $\mathcal{N}=2$ supersymmetry in three dimensions, the associated superspace takes the same form as the one that has $\mathcal{N}=1$ supersymmetry in four dimensions, so one can suspect that the moduli space behaves very similarly: it is always described by local operators in the chiral ring of the theory. However, if we consider a theory with only a single vector multiplet from four dimensions, reduced to three dimensions, the structure of the moduli space looks different from the point of view of perturbation theory. Indeed, a vector particle in three dimensions has only one degree of freedom. The second degree of freedom that would correspond to a second polarization of the vector particle in four dimensions reduces in three dimensions to a real scalar field in the adjoint representation. The scalar field can acquire a vacuum expectation value and generically the unbroken gauge group that is left over after assigning such a vev is given by $U(1)^r$ where $r$ is the rank of the gauge group (we can call these configurations diagonal by analogy with the Cartan elements of $U(N)$ in the fundamental representation). Thus the naive moduli space has dimension $r$ and is not obviously a complex manifold. What happens to restore a complex manifold is that the $U(1)^r$ vector particles that remain massless can be dualized to another set of $r$ real scalar fields that are also massless. These dual scalar fields can acquire vacuum expectation values of their own, and together with the original $r$ diagonal scalars they now form a complex manifold of dimension $r$. Since in order to see all of the degrees of freedom that make up the moduli space we needed to do an electric-magnetic duality in three dimensions, the local  operators that are sensitive to the values of these dual scalars must be constructed non-perturbatively. These local operators that can be used to measure these non-perturbative degrees of freedom are described by magnetic monopole operators. After all, the magnetic monopole objects in three dimensions are instanton-like objects and can be associated to a point (they can be thought of as local  operator insertions). Moreover, they are sensitive to the dual degrees of freedom of the vector field since they couple to those dual degrees of freedom electrically. Hence the vacuum expectation value of such objects can lead to a well defined
set of objects to measure the vacuum structure. However, in pure ${\cal N}=2$ SYM in three dimensions this is a moot point, as the moduli space gets lifted by non-perturbative effects. This story is accurate for   ${\cal N}=4$ SYM and can lead to a full analysis of the moduli space \cite{Seiberg:1996nz}.

Also, unlike four dimensions, in three dimensions there are no anomalies for global symmetries that one could match. Thus a lot of the powerful constraints that in four dimensions allow one to calculate the R-charge are completely absent in three dimensions. However, the AdS/CFT ideas suggest that one can use volume minimization instead to find the R-charge of the theory if one can only compute the associated volume to minimize for the theory.

The story is complicated further because  we can add Chern-Simons terms to the action (this can be done compatibly with supersymmetry \cite{Schwarz:2004yj}), these vector fields generally become massive and can be integrated out of the theory leaving behind an empty moduli space. However, with an appropriate choice of Chern-Simons levels and in the presence of massless matter fields (see for example \cite{Martelli:2008si}), this is not immediate and the Chern-Simons degrees of freedom have consequences for the moduli space of vacua. This is because monopole operators acquire an electric charge due to the Chern-Simons term (see \cite{Borokhov:2002cg}), so the charge can be canceled with matter fields. Such dressed monopole operators would contribute to the chiral ring. The amount of charge that they get dressed with depends on the Chern-Simons level, so the chiral ring structure of these operators is very sensitive to the values of the Chern-Simons couplings and this modifies the moduli space structure. For example, the quantum numbers of dressed monopole operators under global symmetries change when we change the Chern-Simons level \cite{Jafferis:2009th,Benini:2009qs}. This changes the chiral ring relations and if these suffice to characterize the moduli space, then we find different complex varieties for different theories whose only difference in defining their Lagrangians is the assigned value of the Chern-Simons coupling constants.

A second set of problems related to this one is the problem of dualities: different theories can lead to the same infrared physics and thus one can have two different descriptions of the infrared dynamics. Such theories would be called Seiberg-dual \cite{Seibergdual}. However, to match two Seiberg dual theories to each other we need to match the complete chiral ring. This also requires matching the complete spectrum of monopole operators. Understanding when this can be done is still considered an open problem. There also seem to be problems in identifying Seiberg dual pairs \footnote{A brane construction for the case of a single unitary gauge group was done in \cite{Giveon:2008zn,Niarchos:2009aa,Niarchos:2008jb}. For orthogonal gauge groups see \cite{Kapustin:2011gh}. For symplectic gauge groups see \cite{Dolan:2011rp} (in this last paper a proposal connecting the 4d superconformal index with the 3d localized partition function is developed, so, the method can be implemented to find more 3d theories with dual candidates provided it is inherited from a 4d duality). For more complicated gauge groups, a proposal using toric duality is given in \cite{Amariti:2009rb}}. More particularly, such theories require understanding how the different Chern-Simons levels in two different theories are related to each other in order to make the match work. We would also like to compute the R-charge and show that the procedure to compute it is independent of which Seiberg dual frame we choose. This is where the work developed in \cite{Bergman:2001qi,Martelli:2006yb} can work for us: if we can count operators we can make a generating series that counts them (the Hilbert series) and we can expand the answer around the singularity in this series to extract the volume, from which the process of volume minimization can proceed and we can use this as a procedure to calculate the R-charges of all chiral operators.

This paper addresses some of the problems listed above. First, we develop a computational tool that permits us to evaluate the chiral ring semiclassically by considering classical solutions of the equations of motion for a superconformal field theory in the cylinder $S^2\times \mathbb{R}$. The classical solutions saturate the BPS bound where the R-charge is equal to the energy of the configuration. This can be done in the presence of magnetic fluxes. We show that these solutions are covariantly constant on the sphere
and in essence can be described by the matter fields reduced to a matrix quantum mechanical model. We do this assuming a fairly general K\"ahler potential and we show how this information can be bypassed in these classical solutions.
 The solutions are  in correspondence with points in the classical moduli space of the theory as one would have evaluated them in similar theories in four dimensions \cite{Berenstein:2007wi}. Then we show that at the quantum level this set of solutions gets quantized: the set of allowed wave functions are holomorphic on this moduli space.
Also we show that the classical quantization of the magnetic fluxes plays an important role in determining the allowed wave functions of the matter fields and thus on the details of the global topology of moduli space. This is the content of the section \ref{sec:formalism}. We do not consider the further problem of evaluating one loop corrections to the dimension of operators we compute, thus our analysis is valid only when the Chern-Simons levels are large and the ranks are small. This is the weak coupling limit for gauge interactions, but we allow for the possibility of matter being strongly coupled with large anomalous dimensions.

In section \ref{sec:examples} we discuss two examples.  First we study a particular example of theories with $\mathcal{N}=3$ supersymmetry in three dimensions \cite{Jafferis:2008qz}. In this example the complex structure of the system
is free of continuous parameters and instead is quantized by the Chern-Simons levels. We also give an example of studying the theory whose quiver diagram is the same one as that of the $\BC^3/\BZ_3$ quotient \cite{Kachru:1998ys}.

We then go ahead and study possible Seiberg dualities for the ${\cal N}=3$ theories in section \ref{sec:reflections}. We show that these must be realized by Weyl reflections, similar to how these dualities act in four dimensions \cite{Cachazo:2001gh}.
In the special case where the geometry of the $\mathcal{N}=3$ theory has codimension 2 singularities we show that extra monopole operators appear. These
are analogous to fractional branes in orbifold singularities. These give extra branches of moduli space. We show that under some of these Seiberg dualities these fractional brane branches of the moduli space coincide only after quantum corrections to monopole charges are taken into account. Such corrections change the structure of the moduli space. Indeed, in Seiberg-like dualities, at least one of the two theories related by dualities is strongly coupled in the gauge sector.

In section \ref{sec:homol} we discuss a possible systematic realization of Seiberg duality in terms of a change of basis from branes to anti-branes and  bound states of branes, as is usually done in four dimensions for quiver theories \cite{Berenstein:2002fi}. We show that such a candidate duality would lead to inconsistencies for `chiral quiver' theories, where one would be able to change the Chern-Simons levels of the nodes in the quiver repeatedly and that this generally leads to different moduli spaces for such naive dual pairs.

Next, in section \ref{sec:hand}, we discuss how to fix these dualities by hand: we just count monopole operators and match them by comparison between candidate duals. We discuss this in particular for the reduction to three dimensions of the $\mathbb{C}^{3}/\mathbb{Z}_{3}$ quiver gauge theory (from which a particular case is the $M^{1,1,1}$ theory \cite{Fabbri:1999hw}), where we show how some of these dualities that are inherited from four dimensions would appear in three and what are the relations between the Chern-Simons couplings. There are various puzzles that result from this procedure. New theories appear that do not have candidate duals, and other times the conditions under which dualities exist require strange congruence properties of the Chern Simons levels.

We then show that by counting monopole operators in this theory we are able to compute the Hilbert series for these quivers in section \ref{sec:Hilbert}. This is a generating series counting operators of R-charge $a$ with weight $t^a$. The series diverges at $t=1$ and has a pole of fourth order. The R-charge is unknown, but given a hypothetical R-charge we can compute the associated volume of the Sasaki-Einstein space by expanding the series around the pole at $t=1$. Minimizing the volumes with respect to the unknown gives a procedure for calculating the R-charge of the monopole operators and gives a prediction for the volume\footnote{This have been implemented in toric setups, see for example \cite{Hanany:2008fj}.}. The R-charges are generally irrational. This predicts that the dual Sasaki-Einstein spaces are irregular Sasaki-Einstein geometries. We also show that if the Seiberg-like dualities preserve the chiral ring, then the volume minimization procedure gives the operators on both theories the same dimensions.

We then conclude. A review on the the mathematical formalism used for computing solutions to the F-term equations and relevant results are collected in the Appendix.

\section{BPS operators in three dimensional CS matter theories}\label{sec:formalism}

There are two ways of thinking about local operators in a conformal quantum field theory. First, there is an operational point of view where we think of operators as
representations of the conformal group that are to be classified according to their dimension, quantum numbers and algebraic properties under Operator Product Expansions. This point of view is most often used when theories are solvable (like in minimal models in 2D CFT's or in free field theories). There is a second way to view local operator insertions: they can be viewed as an operator that inserts a particular classical singularity in the field configurations at a particular point. This is, it produces a classical solution to the field equations with a particular singular behavior at the operator insertion point \cite{Borokhov:2002ib,Borokhov:2002cg}. At this singularity the equations of motion are not satisfied.

When considering the operator state correspondence, we can also understand that to each operator there corresponds a state in the quantum field theory. These states are defined on the cylinder, so just like operators can be classified in terms of algebraic representations of the conformal group, the same holds for states. The standard way to think of these states in free field theory is in terms of a Fock space of states on a sphere, and the different states are built by considering polynomials in the raising operators for the various spherical harmonics on the sphere. These are in one to one correspondence with polynomials of the fields and its derivatives.

The second class of states would correspond to states that are located very nearby a classical solution to the field equations on the cylinder. These are like coherent states of the quantum field theory and the set of classical solutions needs to be quantized in order to extract the quantum properties of the configurations. This second route to the description of states can reproduce the first one in the case of free field theory. However, the set of classical states can be richer than polynomials in the basic fields of a theory: there can be classical solutions to the equations of motion that are not continuously connected to the trivial solution. These solutions are non-perturbative in nature and their quantization again leads to a description of the set of operators of a theory, but it can include objects that are not polynomials in the fundamental fields. Monopole operators are such states and they are an integral part of three dimensional conformal field theories with Chern Simons terms.

We will describe the set of BPS operators in three dimensional superconformal field theories with $\mathcal{N}=2$ supersymmetry in this fashion. Thus we can go beyond the usual setup where the kinetic term of the fields is canonical and we can consider more general theories in the semiclassical regime. Our general description will follow closely the construction of these states described in \cite{Berenstein:2007wi} for classical BPS solutions on the cylinder in four dimensions. We will adapt those results to three dimensions. These BPS states, properly quantized, give us the chiral ring of the theory. This point of view has also been addressed in studying the ABJM theory \cite{Aharony:2008ug} in \cite{Berenstein:2008dc,Berenstein:2009sa} and for more general operators in \cite{Ezhuthachan:2011kf}. This is also closely related to the Euclidean formulation in \cite{Hosomichi:2008ip}.

As is well known, the chiral ring of a supersymmetric theory serves as a the polynomial ring that describes the moduli space of vacua of the theory as an algebraic variety, and that these varieties are realized as geometries for string compactifications.
 As described in \cite{Berenstein:2007wi}, for many interesting AdS/CFT dual pairs in four dimensional gauge theories, the BPS states that solve the classical equations of motion on a sphere are in one to one correspondence to points in the moduli space of vacua. Their quantization proceeds by realizing that the set of coordinates of the moduli space have an induced symplectic form that is proportional to the K\"ahler form of the associated moduli space variety. Thus one can choose a homolorphic polarization for the  wave functions on the set of BPS classical configurations and this recovers the usual point of view that the chiral ring operators are holomorphic functions on the moduli space of vacua.

The general story for three dimensional gauge theories is more involved. Part of the reason is that
the presence of Chern -Simons terms in the lagrangian restricts these classical solutions in a non-trivial way, by requiring certain classical quantities to be quantized.
One of the main problems of this paper is to address these issues carefully.
We will adapt the tools developed in \cite{Berenstein:2007wi} to our particular case of three dimensional CS matter theories with ${\mathcal N}=2$ Supersymmetry and we will apply it to various examples (many of them have already been addressed in the literature, so this will just recast some of those setups in a new light. We also have  additional results on some of these setups where we can give a more detailed description of the configurations.). Some of our examples will have $\mathcal{N}=3$ supersymmetry, but we will not work exclusively in these setups. $\mathcal{N}=2$ supersymmetry in three dimensions has the same number of supersymmetries as  $\mathcal{\mathcal{N}}=1$ supersymmetry in four dimensions and it can be considered as its dimensional reduction. The multiplet structure can therefore be understood by using the usual superspace techniques of four dimensions (as described in e.g. \cite{Wess:1992cp}) with chiral superfields and vector superfields.

Our main objective is to classify classical solutions of the field theory on a cylinder whose energy is equal to the R-charge. In the quantum theory such states correspond to
short representations of the superconformal algebra. The lowest weight component of such representations is the lowest component of a chiral operator and belongs to the chiral ring (see \cite{Berenstein:2009ay} and references therein for more details).

In what follows we will describe how to solve the equations of motion for a general $\mathcal{N}=2$ theory that saturate the BPS inequality. When properly quantized we will show that we can recover the global topology of the moduli space of vacua of the theory. Since many of these solutions involve monopole operators, one should think of these techniques as being non-perturbative.

\subsection{Classical BPS equations}

We will consider a general superconformal $\mathcal{N}=2$ CS theory plus matter of quiver type, in three dimensions with gauge group $\prod_{i}U(N_{i})$ and bifundamental matter. We will work in the semiclassical regime and we assume that none of the Chern-Simons coupling constants vanish. The action will be described by a K\"ahler potential, a superpotential and the kinetic terms of the gauge fields will be given by a Chern-Simons action (for our examples we  will require it to be of  single trace type for simplicity). We will require that the action be classically invariant under Weyl rescalings. We also have that the R-charge of the chiral fields matches their conformal weight. Apart from the Chern-Simons terms, the arguments are very similar to those in \cite{Berenstein:2007wi}, so some of the full details of proofs can be obtained from that paper.

The supersymmetric CS lagrangian is given by (for a review see \cite{Gaiotto:2007qi})
\begin{eqnarray}
S_{CS}(A_{i})=\frac{1}{4\pi}\int \hbox{Tr}\left(A_{i}dA_{i}+\frac{2}{3}A_{i}^{3}-\overline{\chi}_{i}\chi_{i}+2D_{i}\sigma_{i}\right)
\end{eqnarray}
then we define
\begin{eqnarray}
S^{\mathcal{N}=2}_{CS}=\sum_{i=1}^{r}k_{i}S_{CS}(A_{i})
\end{eqnarray}
where the $k_i$ are coupling constants. They are quantized. The field $\sigma_i$ is a fourth component of the gauge field: it arises from the dimensional reduction of the
degrees of freedom from four dimensions to three.

For the kinetic term of the bifundamental fields we will keep a general form. We will just write the K\"ahler potential
\begin{eqnarray}
S_{kin}=\int d^{4}\theta K(\Phi,\overline{\Phi})
\end{eqnarray}
depending on the chiral superfields $\Phi_{a}$ and with the appropriate inclusions of the gauge multiplet in order to keep it gauge invariant. We will denote $\phi_{a}$ the lowest component of $\Phi_{a}$, with R-charge $\gamma_{a}$. We will consider their scaling dimensions also to be $\gamma_{a}$ (we are looking at superconformal theories after all). We can follow \cite{Berenstein:2007wi} straightforwardly to show that R-charge invariance of $K$ imposes
\begin{eqnarray}
\sum_{a}\gamma_{a}\phi^{a} \partial_{a}K-\gamma_{a}\bar{\phi}^{a} \partial_{\bar{a}}K=0
\end{eqnarray}
and scale invariance $g\rightarrow e^{-2\Omega}g$ gives the condition
\begin{eqnarray}
\sum_{a}\gamma_{a}\phi^{a} \partial_{a}K+\gamma_{a}\bar{\phi}^{a} \partial_{\bar{a}}K=(d-2)K
\end{eqnarray}
where $d$ is the number of dimensions of the theory.
By combining both equations we get
\begin{eqnarray}
\sum_{a}\gamma_{a}\phi^{a} \partial_{a}K=\frac{(d-2)}{2}K\nonumber\\
\sum_{a}\gamma_{a}\bar{\phi}^{a} \partial_{\bar{a}}K=\frac{(d-2)}{2}K
\end{eqnarray}
These can easily be checked for free fields where $K = \bar\Phi \Phi$, and in dimension $d$ we have that $\phi$ has scaling dimension $\gamma= (d-2)/2$. These equations indicate that the K\"ahler potential is a homogeneous function of weight $(a,a)$ with respect to holomorphic/antiholomorphic rescalings, where $2a= d-2$.

If we promote $\Omega$ to a local Weyl transformation (so, it may depend on the coordinates), then in order to have a conformal invariant action, we have to add a term which couples the K\"ahler potential to the Ricci scalar of the metric (this is a non-minimal coupling to gravity for the scalar fields)
\begin{eqnarray}
-\frac{d-2}{4(d-1)}\int d^{d}x \sqrt{-g} K(\Phi,\overline{\Phi})R
\end{eqnarray}

Given the classical field theory, we now write it on the cylinder $\mathbb{R}\times S^{d-1}$. The cylinder can be considered as a Weyl rescaling of flat space (this is radial quantization), where
\begin{equation}
ds^2 = dr^2 +r^2 d\Omega^2 = r^2 \left(\frac{dr^2}{r^2}+d\Omega^2\right)
\end{equation}
If we introduce the radial time $\tau = \log (r)$, the metric is Weyl equivalent to
\begin{equation}
ds^2 \simeq d\tau^2 +d\Omega^2
\end{equation}
and the generator of time translations is the generator of radial rescalings $\partial_\tau = r\partial_r$.
This rescaling establishes the usual operator state correspondence where operators of conformal dimension $\Delta$ are associated to eigenstates of the Hamiltonian on the cylinder with energy $\Delta$. This requires that  the sphere is normalized appropriately: the sphere is a unit sphere, otherwise the conformal dimension of the associated state is equal to the energy on the cylinder up to a rescaling associated to the size of the sphere.

 From the point of view of classical conformal field theory, an operator insertion at the origin is given by a particular
singularity at the operator insertion. This is a point of view heavily emphasized in \cite{Borokhov:2002cg} for the case of three dimensional theories.
This is interpreted as a boundary condition in Euclidean time for radial quantization at $\tau\to -\infty$. Such a boundary condition is interpreted as initial
data when we consider a Lorentzian time. We will study states that saturate the BPS bound: that their conformal dimension is equal to their R-charge.

The solutions that saturate the unitary inequality $\Delta \geq R$ are called BPS. These lead to short representations of the superconformal algebra. The quantization of the space of these classical solutions gives the chiral ring of the theory. This is  the  proposed recipe to compute the chiral ring in \cite{Berenstein:2009ay,Berenstein:2009sa}.

Let us now consider the three dimensional theory on a cylinder $\mathbb{R}\times S^{2}$ with a  unit 2-sphere. The lagrangian is given by
\begin{eqnarray}
S^{\mathcal{N}=2}_{CS}&+&\int dtd\Omega_{2}K_{,a\bar{a}}\left(\mathcal{D}_{0}\phi^{a}\mathcal{D}_{0}\bar{\phi}^{\bar{a}}-\nabla\phi^{a}\nabla\bar{\phi}^{\bar{a}}+[\sigma,\bar\phi^{\bar a }][\sigma,\phi^a] \right)\nonumber\\
&-&\frac{1}{4}\int K-V_{D}-V_{F}
\end{eqnarray}
where we have introduced the comma notation for derivatives in target space $K_{,a}= \partial_{\phi^a} K$, etc and $\mathcal{D}_{0}$ and $\nabla$ denote gauge covariant derivatives in the time direction and the two sphere, respectively.\footnote{ In general we can't make the separation of the term $[\sigma,\bar\phi^{\bar a }][\sigma,\phi^a]$, because, if we have a gauge covariant derivative in an arbitrary K\"ahler manifold, it takes the form $\mathcal{D}_{\mu}\phi^{i}=\partial_{\mu}\phi^{i}-A^{(a)}_{\mu}X^{i(a)}$, where $X^{i(a)}(\phi)$ is a Killing vector on the manifold (see \cite{Wess:1992cp} ch. 24). Therefore $X^{i(a)}$ is not necessarily linear in $\phi^{j}$. For the theories we are considering the matter fields transform linearly under gauge transformations and this part of the lagrangian is correct.} We have explicitly included the contribution of the fourth component of the gauge field $\sigma$. This can be absorbed in the spatial covariant derivative terms if we add a fictitious fourth dimension along which the fields are constant. The field $\sigma$ would be the gauge connection on that direction.

All derivatives are gauge covariant derivatives on the cylinder. $V_{D}$ stands for the D-terms coming from the K\"ahler potential (terms that couple the auxiliary fields of the vector multiplet with the chiral multiplets) and $V_{F}$ are the superpotential terms. The conjugate momenta to the matter fields are given by
\begin{eqnarray}
\Pi_{\phi_{a}}=K_{,a\bar{a}}\mathcal{D}_{0}\bar{\phi}^{\bar{a}}\nonumber\\
\Pi_{\bar{\phi}_{\bar{a}}}=K_{,a\bar{a}}\mathcal{D}_{0}\phi^{a}
\end{eqnarray}
this allows us to compute the classical Hamiltonian and R-charge in terms of the momenta
\begin{eqnarray}
H&=&\int_{S^{2}}\left((K_{,a\bar{a}})^{-1}\Pi_{\phi_{a}}\Pi_{\bar{\phi}_{\bar{a}}}+K_{,a\bar{a}}\nabla\phi^{a}\nabla\bar{\phi}^{\bar{a}}+\frac{1}{4}K+V_{D}+V_{F}\right)\nonumber\\
Q_{R}&=&i\int_{S^{2}}\left(\Pi_{\phi_{a}}\gamma_{a}\phi^{a}-\Pi_{\bar{\phi}^{\bar{a}}}\gamma_{\bar{a}}\bar{\phi}^{\bar{a}}\right)=\frac{i}{2}\int_{S^{2}}\left(K_{,\bar{a}}\mathcal{D}_{0}\bar{\phi}^{\bar{a}}-K_{,a}\mathcal{D}_{0}\phi^{a}\right)
\end{eqnarray}

Notice that in the Hamiltonian we have not added a contribution from the Chern-Simons terms. This is because the Chern-Simons lagrangian is of first order type, and upon making the Legendre transform of the Lagrangian their contributions vanish. This does not mean that they do not contribute to the dynamics.
However, this is the main difference with respect to the analysis of theories in four dimensions. In four dimensions the gauge fields have a Yang-Mills action and they contribute to the energy with the electric and magnetic field squares. The Chern-Simons will contribute to dynamics via the constraints required for gauge invariance of the full theory.

Now, we impose the classical BPS condition $H=Q_{R}$. This equation can be brought to a sum of squares (in the $A_{0}=0$ gauge)
\begin{eqnarray}
\int_{S^{2}}\left((K_{,a\bar{a}})^{-1}(\Pi_{\phi_{a}}+iK_{,a\bar{c}}\bar{\phi}^{\bar{c}}\gamma_{\bar{c}})(\Pi_{\bar{\phi}_{\bar{a}}}-iK_{,c\bar{a}}\bar{\phi}^{c}\gamma_{c})+K_{,a\bar{a}}\nabla\phi^{a}\nabla\bar{\phi}^{\bar{a}}+V_{D}+V_{F}\right)=0\nonumber
\end{eqnarray}
In order for this expression to vanish, all the quadratic terms must vanish separately. This gives us the equations:
\begin{eqnarray}
\dot{\phi}^{a}&=&i\gamma_{a}\phi^{a}, \nonumber\\
\nabla\phi^{a}&=&0\nonumber\\
V_{D}&=&V_{F}=0
\end{eqnarray}
The equation $\nabla \phi^a=0$ includes the spatial derivatives on the sphere $\nabla_{\theta,\varphi}\phi^a=0$ and the covariant derivative in the fictitious dimension $[\sigma,\phi^a]=0$.
From these equations we see that the solutions should be spherically symmetric (their spatial covariant derivative vanishes), so we can take the $\phi^{a}$ fields to be constants on $S^{2}$.

 The first equation above results from using the relations between $\Pi_a$ and $\dot \phi^a$. The resulting differential equation is a first order differential equation (as would be expected for BPS states) and they are trivial to solve. The whole problem reduces to finding initial conditions that satisfy the other equations. These are that the $D$-term and $F$-terms in the potential vanish. Solving the F-terms and D-terms equations up to gauge equivalence is tantamount to saying that the initial condition is a point in the moduli space of vacua.

At this point everything is quite analogous to the four dimensional case. However, in three dimensional CS-matter theories the constraint coming from the equations of motion of $A_{0}$ is different from its 4d counterpart which contains only a SYM action for the gauge fields. These equations establish a relation between the magnetic flux through $S^{2}$ and the charge of the allowed operators that will conform the chiral ring. This equation is given by
\begin{eqnarray}\label{fluxeq}
-\frac{k_{i}F^{(i)}}{\pi}=\int_{S^{2}}-i\sum_{t(a)=i}\Pi_{\phi_{a}}\phi^{a}+i\sum_{h(a)=i}\Pi_{\phi_{a}}\phi^{a}+i\sum_{t(\bar{a})=i}\Pi_{\bar{\phi}_{\bar{a}}}\bar{\phi}^{\bar{a}}-i\sum_{h(\bar{a})=i}\Pi_{\bar{\phi}_{\bar{a}}}\bar{\phi}^{\bar{a}}
\end{eqnarray}
we have denoted by $t(a)$ and $h(a)$  the node where the   tail and head of $\phi_{a}$ end respectively. These should be interpreted as matrix equations in $U(N)$, but their meaning is very clear in the case that the theory only has a $U(1)$ gauge field: the magnetic flux induces electric charge, and the restrictions for gauge invariance require
the net electric charge to vanish, with the matter canceling exactly the charge carried by the magnetic flux.

It is easy to show that the $F^{(i)}$ are also covariantly constant (and in the gauge where $\phi$ is constant, $F$ is also constant). The magnetic flux is real and it is described by a self-adjoint matrix that can always be diagonalized. This breaks the $U(N)$ gauge group to a product of $U(N_i)$ with $\sum N_i=N$.

This should be considered as a dimensional reduction to a matrix quantum mechanics. This is a finite dimensional phase space with matrix degrees of freedom $\phi$ and $\Pi$. The Chern-Simons lagrangian does not carry a net number of degrees of freedom (it is topological). Since we consider only spherically symmetric configurations we have that the electric fields vanish and that the magnetic field strength is constant on the sphere. This needs to be included in the description.

Notice moreover that the magnetic flux is quantized for $SU(N)$ so the flux equation restricts the solutions of the F-terms and D-terms in a non-trivial way.
The quantization conditions for $U(1)$ can be subtle as this dictates the possible baryonic charges \cite{Benishti:2010jn} (for an example see \cite{Berenstein:2009sa}).
This concludes the computation of classical BPS equations on the cylinder for superconformal CS-matter theories. We will now analyze the solutions to these equations in detail, as well as their quantization.

\subsection{Holomorphic quantization of the moduli space}

So far, we can solve some of the BPS equations in a straightforward manner: we require the scalar fields to be constant on the sphere and we reduce to a matrix model
with finitely many matrices. Now, there are various routes to proceed: we can first quantize this set of configurations and then impose the rest of the BPS equations  as constraints on the wavefunctions that we allow. We can also solve for points in the moduli space of vacua and quantize these configurations. Finally, we can take a mixed approach where some things are solved for first and quantization occurs at this intermediate step, while solving some other constraints later on. This is what we will do, and we will ignore the subtleties associated to possible quantum corrections, assuming that supersymmetry will solve them for us at the end of the day. The procedure we outline is valid in the semiclassical approach. We will solve the first order equations of motion $\dot\phi^a=i\gamma_a \phi^a$. These configurations do not automatically live in the moduli space of vacua, so the other BPS equations from the F-term and D-terms have not been imposed yet.

Before imposing these other BPS constraints, the symplectic form on the classical phase space, $\mathcal{M}$,  is just
\begin{eqnarray}
\omega=d\Pi_{\phi_{a}}\wedge d\phi^{a}+d\Pi_{\bar{\phi}_{\bar{a}}}\wedge d\bar{\phi}^{\bar{a}}
\end{eqnarray}
The traces over matrix degrees of freedom are implicit.

By solving the constraints associated to the first order equations we get
\begin{eqnarray}
\Pi_{\phi_{a}}&=&-i\sum_{\bar{a}}K_{,a\bar{a}}\gamma_{\bar{a}}\bar{\phi}^{\bar{a}}=-\frac{i}{2}K_{,a}\nonumber\\
\Pi_{\bar{\phi}_{\bar{a}}}&=&i\sum_{a}K_{,a\bar{a}}\gamma_{a}\phi^{a}=\frac{i}{2}K_{,\bar{a}}
\end{eqnarray}

We notice that the first order equations tell us that we can use either $\Pi_{\phi^a}$ and $\phi^a$ as the canonical  variables of the reduced phase space that solves the constraints (as the K\"ahler metric is invertible, except perhaps on a set of measure zero), or we can trade those by $\bar \phi^{\bar a}$ and $\phi^a$.

If we choose the second set of coordinates, we find that
on this submanifold of BPS states that the
 the pullback of $\omega$ is given by
\begin{eqnarray}
\omega=iK_{,a\bar{a}}d\phi^{a}\wedge d\bar{\phi}^{\bar{a}}=-2d\phi^{a}\wedge d\Pi_{\phi^{a}}
\end{eqnarray}

We get exactly the K\"ahler form of the metric. The scaling property of the metric is crucial for this.

We see that the classical BPS states can be associated to a phase space that is a K\"ahler manifold itself.  We can holomorphically quantize the space of classical solutions by choosing a holomorphic polarization. Indeed, we see from $\omega$ that the variables $\phi^a$ have vanishing Poisson brackets, so we can `diagonalize' them simultaneously as quantum variables (not as matrices) and write the wave functions as functions of the $\phi^a$, but not their canonical conjugates.
Choosing this representation, the wave functions will be functions only of the $\phi^{a}$'s and therefore are holomorphic. Regularity at the origin will impose that the wave functions are polynomial. To add the F-term constraints we can impose the F-term relations as relations between the $\phi$ variables. Hence the wave functions can have many representatives, where they differ by F-terms. This is getting closer to the usual notion of the chiral ring. Notice that because the holomorphic variables commute with each other under Poisson brackets, there is no ordering ambiguity and the symbols can be manipulated as if they are numbers or regular variables.

Under these conditions, we have the standard representation $\Pi_{\phi_{a}}\sim \partial_{a}$ and the wave functions we will consider will be taken to be polynomials in the $\phi_{a}$'s modulo F-terms. This is usually stated by saying that the wave function is single-valued and regular at the origin.

The flux operator appearing in (\ref{fluxeq}) for the case of a $U(1)$ field  becomes equal  to a number operator that counts the number of arrows that leaves a node minus the arrows that enter in that node.  The restriction that the flux is integer on the left hand side puts a restriction on the allowed polynomials of the fields $\phi$. The arrows are counted by the degrees of  the fields $\phi$.
To be explicit, if our holomorphic wave functions are monomials of the form
\begin{eqnarray}
\psi=\prod_{a}\phi_{a}^{m_{a}}
\end{eqnarray}
the equations (\ref{fluxeq}) can be written as a differential equation where the left hand side is interpreted as a number (or set of numbers), and the right hand side is interpreted as a first order differential operator. Hence
\begin{equation}
-{k_{i}F^{(i)}}\psi = \left(-\sum_{t(a)=i} \phi^a \partial_a +\sum_{h(a)=i} \phi^a\partial_ a\right)\psi\label{eq:fluxtwo}
\end{equation}
which upon substitution becomes
a simple constraint on the exponents
\begin{equation}
-k_{i}F^{(i)}=-\sum_{t(a)=i}m_{a}+\sum_{h(a)=i}m_{a}
\end{equation}
Notice also that this equation is compatible with the F-term constraints, as those are gauge covariant and preserve the charge of the polynomials in the $\phi$ as well as the R-charge.

Classically for spherically invariant configurations the fluxes $F^{(i)}$ are covariantly constant and can be made diagonal by a gauge transformation. A non-zero value of $F^{(i)}$ corresponds to what are called monopole operators. These are also automatically classical solutions of the Yang-Mills theory on the 2-sphere and solve not only the Chern-Simons equations of motion, but also the Yang-Mills + Chern-Simons equations of motion in the presence of matter. We will not need this for this paper. The presence of the gauge flux changes the topology of the matter fields (for example, matter fluctuations can be quantized according to monopole spherical harmonics rather than regular spherical harmonics \cite{Borokhov:2002ib,Borokhov:2002cg}). However, when we consider these classical solutions we have the consistency condition
\begin{equation}\label{sphersym}
[\nabla_\theta, \nabla_\varphi] \phi^a =0 = [F_{\theta\varphi}, \phi^a]
\end{equation}
This implies that the field $\phi^a$ commutes with the flux $F$: a non zero vev for $\phi^a$ can only connect two gauge fluxes that have the same numerical value. We can therefore split the solutions into block-diagonal sectors according to a single integer: the flux for an eigenvalue of $F$ (i.e. the flux in each gauge group). The trace of the flux is invariant: it is the first Chern class of a bundle on a Riemann surface (the sphere). These are conserved not just for BPS states but for all states. Hence they should be associated to conserved charges in the dual string theory/M-theory setup and to an associated gauge field in the AdS theory.

For a more general $U(N)$ theory, we need to write these as matrix equations, and if the flux term is proportional to the unit matrix the equations tell us that the polynomial of the $\phi^a$ is $SU(N)$ invariant for each gauge group, while at the same time carrying the right kind of charge under the diagonal $U(1)$. In other setups we think of the left hand side of eq. (\ref{fluxeq}) (the $F$) as being quantized by classical topology and thus each set of $F$ defines a superselection sector. Since it can be diagonalized by classical gauge transformations we will assume that the left hand side is always a diagonal matrix. We write these as block diagonal form where the blocks have common flux eigenvalues.
We will write some examples later in the paper where we consider some setups that require $U(N)$ fields rather than products of $U(1)$'s. For the present discussion we assume implicitly that this is  part of the equations above.

The equations above can be interpreted as eigenvalue equations for the non-zero $F$ in the different gauge groups $U^{(i)}$: we can count the amount of charge that flows into a given eigenvalue of $F$ by counting arrows with the right gauge index structure. Since arrows that connect different eigenvalue blocks of $F$ carry angular momentum (they are quantized according to monopole spherical harmonics), they will have zero value classically and in the quantum theory we set them to the ground state of the corresponding oscillators, hence they will not contribute to the description of the operators \footnote{Equivalently, for vacuum configurations, all the fluxes $F^{(i)}$ on the different vertices of the quiver are equal.}. This is, we will not include them in the reduction of phase space to the matrix model of constant fields.
All solutions to the equations are going to be block-diagonal, where the block are characterized by common eigenvalues of the flux.
One can also check that the D-term equations are very closely related to the flux equations.

Incidentally, if one works in detail the D-term equations and we impose them for BPS states, we find that $\sigma \propto F$, so requiring that the gauge field commutes with the matter configuration automatically guarantees that $\sigma$ will commute with the matter configuration and we do not have to do extra work to solve this equation. This is, the equation of motion for $\sigma$ will be the same equation that determines the flux given the matter.

As an aside, realizing that at large $k$ (the perturbative string theory regime) these states cost an  energy of order $k$, since after all the energy is given by the R-charge of the polynomial of the $\phi$, we find that the states that carry this charge should be interpreted as D-branes, and the conserved charge should be paired to some Ramond-Ramond field wrapping a cycle of the compactification.

We will now apply this set of tools  to a variety of different examples.

 Here, we use the ideas advocated in \cite{Berenstein:2000ux,Berenstein:2001jr,Berenstein:2002ge}: the F-term equations are interpreted as defining an algebra.
  The solutions of the F-term equations are representations of the said algebra and they can generically be described
by direct sums of irreducible representations. Reducible representations that are not direct sums can also show up. These depend on the D-terms and generally lead to
a bigger object: a category of modules over the algebra. This also leads to  a more general category of holomorphic branes (the derived category of modules of the algebra
\cite{Sharpe:1999qz,Douglas:2000gi}) which will play a role when we describe Seiberg dualities. This is elaborated with more detail in the appendix where we give a guide to some of the literature and basic definitions governing these algebras.

\section{Two examples}\label{sec:examples}

\subsection{A first example: the $A_2$ theory}

The simplest theory that one can analyze is the ABJM model \cite{Aharony:2008ug,Aharony:2008gk}. In that case the gauge group is $U(N)\times U(N)$, and a single brane in the bulk corresponds to a $U(1)\times U(1)$ theory. For such a setup the superpotential vanishes after integrating out the auxiliary fields of the vector multiplet and the theory is essentially free, except for the quantization condition of monopole operators. The description of the moduli space and quantization of classical solutions with the techniques that we have developed so far was done in detail in \cite{Berenstein:2008dc,Berenstein:2009sa}, in that setup since the theory could be described essentially in free field theory the holomorphic quantization was fairly trivial.

For our first non-trivial example we will consider a theory with $\mathcal{N}=3$ supersymmetry in three dimensions that is similar to the ABJM model \footnote{These families of theories were proposed in \cite{Jafferis:2008qz}.}, but where there are F-term constraints that need to be satisfied. We will set up the theory associated to the $A_2$ singularity in four dimensions, dimensionally reduce and add the $\mathcal{N}=3$ Chern-Simons lagrangian for the vector multiplet. The theory is best described by the extended Dynkin diagram of $SU(3)$. This is depicted in figure \ref{fig:A2quiver}.

\FIGURE{
\includegraphics[width=5cm]{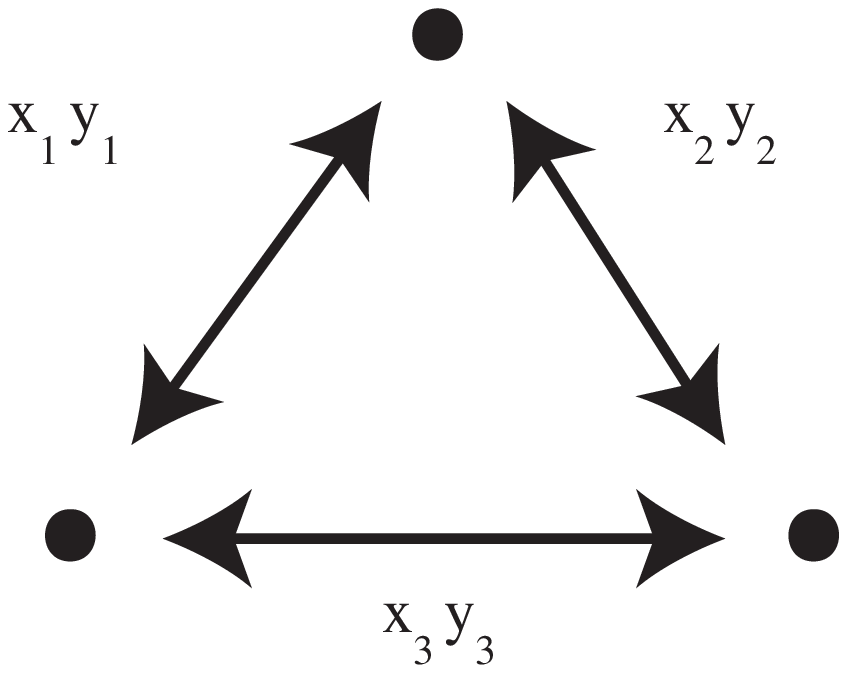}
\caption{Figure of the $A_2$ quiver theory}\label{fig:A2quiver}}

The theory consists of three vector multiplets ${\bf V}_1,{\bf V}_2,{\bf V}_3$ for gauge groups $U(N_1)$, $U(N_2)$, $U(N_3)$ respectively,  and three hypermultiplets, ${\bf X}_1,{\bf X}_2,{\bf X}_3$. These transform in the $(N_i, \bar N_{i+1})$, where $N_{i+3}\simeq N_i$. We will split these according to the $\mathcal{N}=1$ superspace notation in four dimensions.
${\bf X}_i$ will consist of two chiral superfields $X_i$ and $Y_i$, where $Y_i$ transforms in the conjugate representation of $X_i$: $\bar Y_i$ is the antiholomorphic superpartner of $X_i$. Similarly the vector superfield ${\bf V}_i$ will split into an $\mathcal{N}=1$ vector superfield $V_i$ and a holomorphic partner $Z_i$. The lagrangian is determined uniquely by the choice of Chern-Simons couplings $k_{1,2,3}$ and the $\mathcal{N}=3$ supersymmetry. The $X,Y$ superfields have canonical kinetic terms.
 For us, the most important piece of knowledge is the superpotential, which is given by
 \begin{equation}
 W_{A_{2}} = \sum_{i=1}^3 \tr ( X_iY_i Z_{i+1} - Y_iX_i Z_i -\frac 12 k_i Z_i^2)\label{eq:A2quiverW}
 \end{equation}
 In this setup both $X,Y$ have canonical dimension $1/2$ in 3 dimensions, but $Z$ has dimension $1$. This theory has been analyzed in detail in \cite{Jafferis:2008qz}

In the conformal theory the fields $Z_i$ do not have a kinetic term and can be integrated out. It is more convenient however to keep them in the superpotential for computation of the F-terms. We will be interested in the case where $\sum k_i=0$. This superpotential is a special case of those considered in \cite{Cachazo:2001gh,Berenstein:2002ge}.
If this where a four dimensional theory we would obtain a moduli space given by a symmetric product of copies of the generalized conifold \cite{GNS}
\begin{equation}
uv = w (w+k_1 z ) (w+(k_1+k_2) z)\label{eq:chiral}
\end{equation}
where the gauge group is $U(1)^3$ and $u= x_1x_2x_3$, $v=y_1y_2y_3$ and $z$ is any of $z_1,z_2,z_3$. The F-term relations make them all equal to each other.
Here the associated dimension of $u,v$ is $3/2$ and for $w, z$ the associated dimension is equal to $1$.

We should notice that from the point of view of complex geometry it is interesting to notice that the complex structure of the variety associated to equation \eqref{eq:chiral}
is quantized: the complex structure is determined entirely by integers. In dual AdS theories this gives a simple example that shows that some moduli are stabilized by fluxes.

Alternatively, we can think of matrices
\begin{eqnarray}
X&=& \begin{pmatrix} 0 & X_1 & 0\\
0&0& X^2\\
X^3&0&0
\end{pmatrix}, \quad Y= \begin{pmatrix} 0 & 0 & Y_3\\
Y_1&0& 0\\
0&Y_2&0
\end{pmatrix}, \quad Z= \begin{pmatrix} Z_1 &0&0\\
0&Z_2&0\\
0&0& Z_3\end{pmatrix}\\
K&=& \begin{pmatrix} k_1 &0&0\\
0&k_2&0\\
0&0& k_3\end{pmatrix}
\end{eqnarray}
and all the F-term relations can be encoded in the following algebraic equations
\begin{eqnarray}
[Z,X] &=& [Z,Y]= [Z,K] =0\\
{}[X,Y] &=&- K Z
\end{eqnarray}
These can be considered as equations determining an algebra. The solution of these equations is a representation of the algebra, which generically can be decomposed into direct sums of irreducibles. The techniques developed in
\cite{Berenstein:2000ux,Berenstein:2001jr,Berenstein:2002ge} show  that the moduli space of  four dimensional theories (solution of the F-terms) is in general a symmetric product because of this decomposition (for a classification of representations of this algebra and $A_{n-1}$ in general, see \cite{Cachazo:2001gh,Berenstein:2002ge}).

The first line shows that $Z$ should be considered to be in the center of the algebra. Similarly one can show that $X^3=U$ is in the center, as well as $V=Y^3$.
From here it follows that
\begin{equation}
UV = (W)(W+k_1 Z)(W+(k_1+k_2) Z)
\end{equation}
where $W$ is a fourth variable determined in \cite{Berenstein:2002ge}. Numerically it is given by $x_3y_3$ on a single brane setup (where the gauge group is $U(1)^3$). The other two
combinations $W+k_1 Z$ and $W+(k_1+k_2) Z$ are given by $(x_1y_1)$ and $(x_2y_2)$ respectively.

 This equation is true at the level of the algebra of F-terms and it determines the center of the algebra. The geometry is singular at the origin. It is also singular if some of the roots of the polynomial on the right hand side are repeated. This would imply that at least one of the three $k_i$ is equal to zero. This situation can not be analyzed in perturbation theory nor classically, as $1/k_i$ play the role of perturbative coupling constants. So the situation where one $k_i=0$ is highly singular and can not be analyzed with our current techniques, except by extrapolation.

The so called mesonic elements of the chiral ring are traces of polynomials of $U,V,W,Z$, subject to the $F$-term constraints. For the $U(1)^3$ theory the chiral ring
relation are exactly those of \eqref{eq:chiral}.
This particular theory has also been discussed in \cite{Jafferis:2008qz} \footnote{ The relations would correspond to those of eq. (2.28) in their work. However they have mistyped the correct solution found in \cite{GNS,Berenstein:2002ge}. }

Remember that so far in our construction we have that the variables $X_i,Y_i$ are holomorphic coordinates. So are then $U,V,W,Z$. These are well defined coordinates on the moduli space of vacua and according to our quantization prescription they automatically become associated to states in the chiral ring.  The requirement of taking traces is to ensure gauge invariance of the configuration.

Our purpose is not to re-derive these results, but to instead consider the spectrum of monopole operators of the theory. So we need to add gauge flux and consider what type of states we are allowed to have.  As discussed in the first section, on an irreducible we have that the flux commutes with the configuration, so we need to add the additional generator for flux, let us call it $F$. The equations for $F$ are given by
\begin{equation}
[F,X]=[F,Y]= [F,Z]= 0
\end{equation}
and $F^\dagger = F \propto \sigma$. The new hermiticity constraint makes $F$ real and turns the algebra into a $C^*$ algebra. When we solve for $X,Y, Z$, we also need to solve for the D-term equations. This is assumed throughout, but it is somewhat involved to get exact matrices that satisfy the D-term relations. Because we are in the end considering a holomorphic quantization this is not strictly necessary, as the chiral ring will be only determined by polynomials of holomorphic variables.

Let us consider the simplest set of monopole operators: those with $F=\pm 1$. We need to solve the flux equation given by
\begin{equation}
\pm\begin{pmatrix}
k_1 &0&0\\
0&k_2&0\\
0&0& k_3
\end{pmatrix} \psi = \begin{pmatrix} x_1 \partial_1 + y_3 \partial'_3 &- x_3 \partial_3- y_1\partial'_1 &&\\
& x_2 \partial_2 + y_1 \partial'_1 &- x_1 \partial_1- y_2\partial'_2 & \\
&& x_3\partial_3 + y_2 \partial'_2 &- x_2 \partial_2- y_3\partial'_3
\end{pmatrix} \psi \label{eq:flux}
\end{equation}
In this equation derivatives with respect to the $x_i$ are denoted just by $\partial_i$. derivatives with respect to the $y_i$ are denoted by $\partial'_i$. Here we use lower case to indicate that the $X_i$ are $1\times 1$ matrices, and thus can be considered as scalars.

It is easy to see that any polynomial of traces of $U,V,Z,W$  is killed by the differential operators on the right hand side. Thus any solution of the monopole equations for $\psi$ can be multiplied by traces of $U,V,Z,W$ and this generates a new solution. What we want to find is those solutions that are minimal.Let us assume for simplicity that $k_1, k_2>0$ and that $k_3<0$. After all, two of the $k_i$ have to have the same sign, and the cyclic symmetry of the quiver lets us set it up so that we can make this choice.
It is easy to check that $y_3^{k_1} x_2^{k_2}$ is an allowed polynomial for $\psi$ when $F=1$. We can also exchange $y_3$ by $x_1 x_2$, or exchange $x_2$ by $y_3y_1$. If we do both exchanges we get terms including powers of  $x_1 y_1$. These can be stripped, as they have already showed up in polynomials of $U,V,W,Z$ and are not new. Thus we get a list of elements given by
\begin{eqnarray}
&y_3^{k_1} x_2^{k_2}& \\
& y_3^{k_1-s} x_2^{k_2} (x_1 x_2)^s & \quad \hbox{ For $s=1,\dots, k_1$}\\
&y_3^{k_1} x_2^{k_2-s} (y_1 y_3)^s &\quad \hbox{ For $s=1,\dots, k_2$}
\end{eqnarray}
Let us call these $P, S_{1, \dots, k_1}, T_{1, \dots k_2}$. These can be conveniently thought of in terms of decorating the quiver with labels that indicate how many arrows
are turned on from which node and in which direction.

\FIGURE{
\includegraphics[width=5cm]{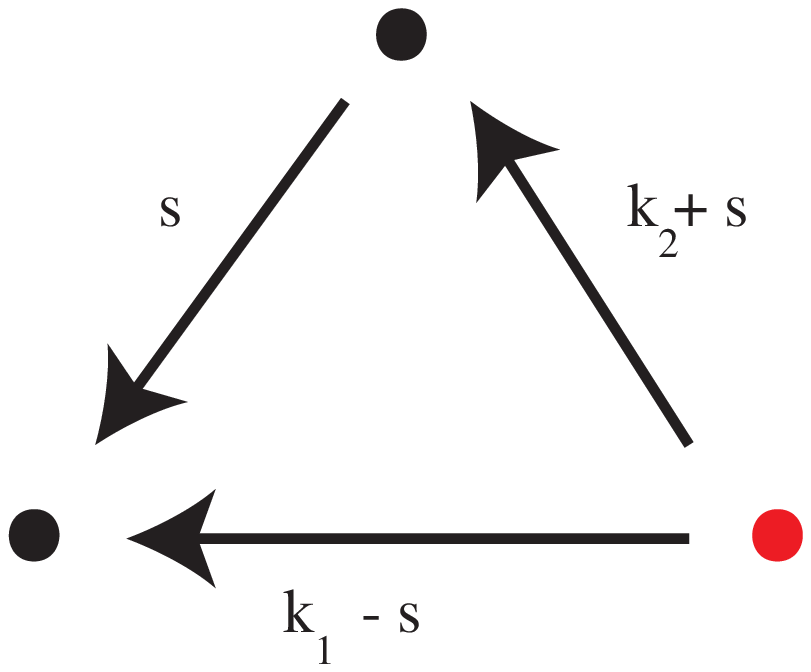}
\caption{Figure of the decorated $A_2$ quiver theory corresponding to the wave function $y_3^{k_1-s} x_2^{k_2+s} x_1^s$. Node three is decorated as source and the other nodes as sinks. }\label{fig:A2monopoles}}

An example is shown in figure \ref{fig:A2monopoles}

Similarly, when we make $F=-1$, we get that the simplest generator is $x_3^{k_1} y_2^{k_2}$ and similarly we can replace  $x_3$ by $y_1 y_2$, or $y_2$ by $x_1x_3$. We thus get another list of generators given by $\tilde P, \tilde S_{1, \dots, k_1}, \tilde T_{1, \dots k_2}$. The decorations just reverse the arrows in the quiver and the role of sources versus sinks.

The relations between these are for example $P \tilde P \simeq h (W,Z)$, $P\tilde S_i \simeq V^i f_i(W,Z)$, $P\tilde T_i \simeq U^i g_i (W,Z)$ where $h, g, f$ are polynomials of only $W,Z$. Similar relations hold for products $S_i \tilde S_j$ etc. These equations tell us that given $W, Z, U, V, P$, we can determine all the other variables uniquely.
$P$, the simplest monopole wave function is unconstrained, and there is a relation between the variables $W,Z,U,V$. Thus we have that the variety describing the allowed polynomials of the variables is four dimensional. Also notice that since $P \tilde P \simeq h (W,Z)$, we can think of $P, \tilde P$ as determining a $\BC^*$ bundle over the base $W,Z,U,V$ which degenerates on $h=0$. The polynomial $h$ is easy to determine. It is given by $W^{k_1} (W-k_3 Z)^{k_2}$.

It is easy to see that given all these variables we can solve the general flux equation for different fluxes, by taking powers. Indeed, for $F=\pm \ell$, the polynomials $P^\ell$ and $\tilde P ^\ell$ have the minimum number of letters and the same substitution rules that we have before apply.

Notice also that if $k_1, k_2$ are not coprime, this is, if $k_1= k_1' \ell$ and $k_2 = k_2' \ell$, then the chiral ring of the theory has many of the same elements of those of the theory for $k_1', k_2'$. It is exactly those where the power of $P$ or $\tilde P$ is a multiple of $\ell$, or more generally, if we define the $\ell$-ality of $P, S,T$ as $+1$, and the $\ell$-ality of $\tilde P, \tilde S, \tilde T$ as $-1$, the allowed products of $P, S,T, \tilde P, \tilde S, \tilde T$ are those that have $\ell$-ality equal to $\ell$ (we take the $\ell$-ality to be additive). This is the orbifold of the variety given by $V_{k_1,k_2}$ by the $\BZ_\ell$ that acts according to
\begin{eqnarray}
(P,S,T) &\to& \exp (2\pi i /\ell) (P,S,T)\\
(\tilde P,\tilde S,\tilde T) &\to& \exp (-2\pi i /\ell) (\tilde P,\tilde S,\tilde T)\\
(U,V,W,Z) &\to & (U,V,W,Z)
\end{eqnarray}
This is very similar to how in the ABJM model the Chern-Simons level gives an orbifold space. Here we see the same pattern explicitly at the level of the description of the monopole operators, this was also noticed in
\cite{Martelli:2008si}. The first appearance of this phenomenon for M-theory moduli space is in the ABJM model \cite{Aharony:2008ug}.

Obviously the variety described here is rather complicated in terms of the generators and relations. The relations can not be described as a product of monomials, so the variety is not toric. Also, the monopole wave functions have a complicated product structure of constraints. Obviously, it is easier to describe this space in terms of the coset geometry  $U(1)\backslash U(3)/U(1)$ as done in \cite{Jafferis:2008qz}. For other $\mathcal{N}=3$ theories the hyperk\"ahler quotient description is more complicated. Our techniques work very similarly there.

Notice that since the theory has $\mathcal{N}=3$ supersymmetry, the dimensions of all operators are determined by the R-charge, which is part of an $SU(2)$. We just need to count the number of $x, y$ to determine the dimension.

\subsection{A second example: the $\BC^3/\BZ_3$ quiver. }\label{sec:c3z3}

Our second example again starts from a four dimensional theory, which is associated to the $\BC^3/\BZ_3$ orbifold. The theory has
$3$ vector superfields $V_{1,2,3}$ corresponding to gauge groups $U(N_i)$. There are also $9$ matter fields $X_{1,2,3}, Y_{1,2,3},Z_{1,2.3}$. The $W_i$ transform as a $(N_i, \bar N_{i+1})$, again with $N_{i+3}\simeq N_i$. The superpotential is given by
\begin{equation}
\tr( XYZ - XZY)
\end{equation}
where we have that
\begin{equation}
X= \begin{pmatrix} 0& X_1&0\\
0&0&X_2\\
X_3&0&0\end{pmatrix}, \quad Y= \begin{pmatrix} 0&Y_1&0\\
0&0&Y_2\\
Y_3&0&0\end{pmatrix}, \quad Z= \begin{pmatrix} 0& Z_1&0\\
0&0&Z_2\\
Z_3&0&0\end{pmatrix}
\end{equation}

 We dimensionally reduce this theory to three dimensions and add Chern-Simons terms to the theory by considering levels  $k_1, k_2,k_3$. Again, we require that the levels add up to zero $k_1+k_2+k_3=0$, and for simplicity we assume that $k_1, k_2 >0$. The quiver diagram appears in figure \ref{fig:C3-Z3}.

\FIGURE{
\includegraphics[width=5cm]{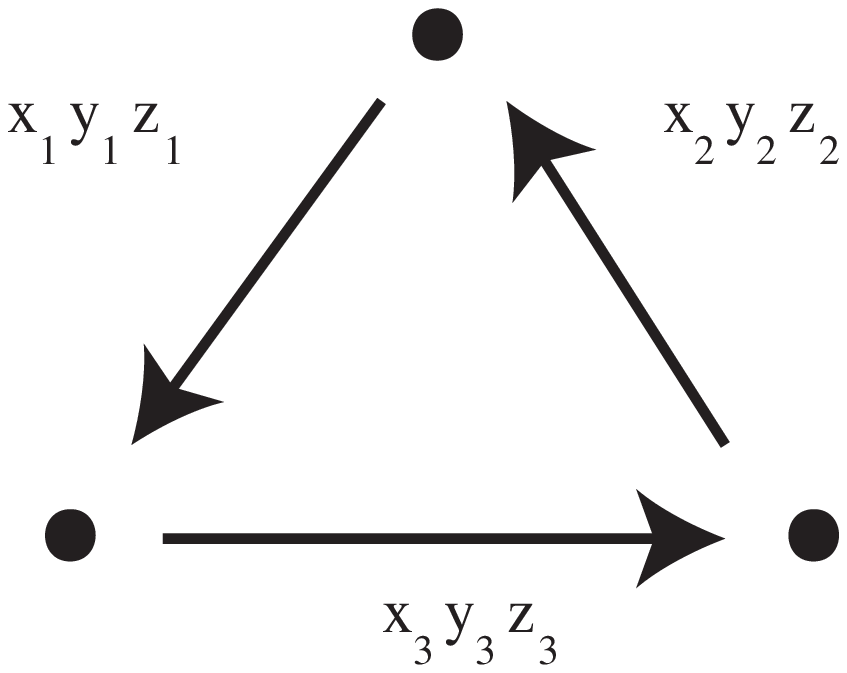}
\caption{Figure of the $\BC^3/\BZ_3$ quiver theory}\label{fig:C3-Z3}}

 The theory has a manifest $SU(3)$ flavor global symmetry and the $(X_i,Y_i,Z_i)$ form a $\mathbf{3}$ of $SU(3)$. The fields $X_i,Y_i,Z_i$ all have the same R-charge, but their
 individual R-charges are undetermined as of yet from field theory considerations.

 Again, the F-terms that follow from the superpotential above give that $[X,Y]=[Y,Z]= [Z,Y]=0$. However, $X,Y,Z$ do not belong to the center, as they do not commute
 with the projections that define the nodes of the quiver (and define the $U(1)^3$ gauge transformations). However, any cubic term made of $X,Y,Z$ does, and these are the ones that can be used to define the mesonic branch of the quiver by taking traces. One gets this way ten different variables with various relations. These are described explicitly in \cite{Berenstein:2002ge}, but can also be described in terms of toric geometry (see for example \cite{Davey:2009qx}). It is easy to show that the irreducibles always have a $U(1)^3$ gauge symmetry and that the associated Calabi-Yau geometry has an isolated singularity in codimension 2.

 The main observation of \cite{Berenstein:2002ge} is that the mesonic operators that generates the center of the algebra can be organized in a single $\mathbf{10}$ of $SU(3)$
 with three totally symmetric boxes in their Young tableaux. Let us call these $U_\alpha$, $\alpha=1,\dots 10$. The relations are such that in the products only totally symmetric representations with $3N$ boxes appear.
 This will aid us in counting operators later in the paper.

 What we want to do now is to consider the monopole operators. Again, we will get equations similar to those found in \eqref{eq:flux}. We will write these schematically as follows
 \begin{equation}
 F \begin{pmatrix} k_1 &0&0\\
 0 &k_2&0\\
 0 &0&k_3
 \end{pmatrix} \psi = \begin{pmatrix} w^3\partial_3-w^1\partial_1 &0&0\\
 0 &w^1\partial_1-w^2\partial_2&0\\
 0 &0&w^2\partial_2-w^3\partial_3
 \end{pmatrix}
 \end{equation}
 where $w^i \partial_i = x_i \partial_{x_i}+ y_i\partial_{y_i}+z_i\partial_{z_i}$ is a place holder for any of $x,y,z$ summed over. The same reasoning than in the previous section shows us that the operator on the right acts by zero on traces of the $U_\alpha$, so any solution to the equation can be multiplied by polynomials of the $U_\alpha$ to obtain new solutions. Those that do not have products forming triangles are the allowed operators.

When $F=1$,  they are schematically of the form
\begin{equation}
(w^3)^{k_1+k_2} (w^1)^{k_2}. \label{eq:MonoC3}
\end{equation} Let us call these $M_\alpha$. Again, the $w^i$ can be replaced by any of $x^i, y^i, z^i$. Coupled to the F-term relations we find that these wave functions transform as a totally symmetric representation with $2k_1 +k_2$ boxes. If we consider positive integer flux $\ell$, we get generically that the allowed wavefunctions are products of the ones with $\ell=1$, and they are of the form  $(w^3)^{\ell(k_1+k_2)} (w^1)^{\ell(k_2)}$.

Similarly, with $F=-1$, we get wavefunctions of the form $(w^2)^{k_1+k_2} (w^1)^{k_1}$, which we will call $\tilde M_\alpha$.
The relations between these are of the form
\begin{equation}
M_\alpha \tilde M_\beta \sim (U_\gamma)^{k_1+k_2}\label{eq:Z3rels}
\end{equation}
 which again show that once the $U_{\alpha}$'s are known (which determine a three dimensional geometry) and one of the $M$ variables is known, the other $M$ and $\tilde M$ variables are determined by the relations. This proves that the associated variety is a four dimensional geometry. Similar considerations to those on the previous section also show that if $k_1,k_2$ are not coprime, but instead $k_1= s k_1'$, $k_2=s k_2'$ then the total moduli space variety is a $\BZ_s$ orbifold of the theory associated to the Chern-Simons couplings $k_1', k_2'$, and that choosing one of each $S,T$, one has a in general a $\mathbb{C}^*$ bundle over the base $\BC^3/\BZ_3$.

\section{Seiberg-like dualities for $A_{n-1}$ quivers}\label{sec:reflections}

We will now consider more general $A_{n-1}$ quiver diagrams with $\mathcal{N}=3$ supersymmetry \cite{Jafferis:2008qz}. These are described by a cyclic quiver diagram with $n$ nodes, as depicted in figure \ref{fig:A_Nquiver}.

\FIGURE{
\includegraphics[width=7cm]{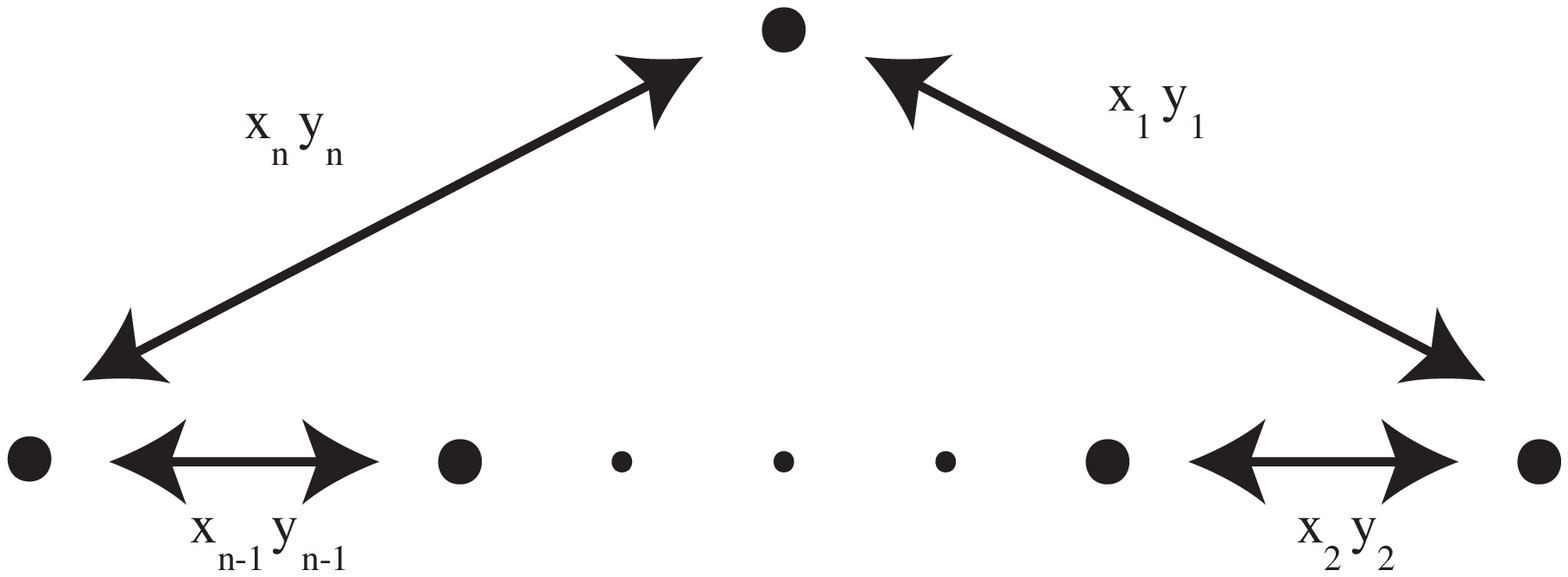}
\caption{Figure of the $A_{n-1}$ quiver theory}\label{fig:A_Nquiver}}

The theory again results from taking supersymmetric theories in four dimensions with $\mathcal{N}=2$ supersymmetries associated to $A_{n-1}$ quiver diagram, reducing to three dimensions and  adding Chern-Simons terms.

The superpotential is identical to the one appearing in equation \ref{eq:A2quiverW}, except that the sum runs all the way up to $n$
 \begin{equation}
 W_{A_{n-1}} = \sum_{i=1}^n \tr ( X_iY_i Z_{i+1} - Y_iX_i Z_i -\frac 12 k_i Z_i^2)\label{eq:ANquiverW}
 \end{equation}
 Also, the solution to the F-terms is entirely analogous. We can define matrices $X,Y,K$ so that the superpotential takes the form
 \begin{equation}
W_{A_{n-1}}= \tr ( [X,Y] Z - \frac 12 K Z^2)
 \end{equation}
Again, the $k_i$ are integers and we will also require that $\sum k_i=0$. Again, following \cite{GNS,Berenstein:2002ge}, we obtain an algebra where there are four variables and one constraint:
\begin{equation}
UV= W( W+k_1 Z)(W+(k_1+k_2) Z) (W+(k_1+k_2+k_3) Z) \dots (W-k_{n} Z) =C(W,Z)\label{eq:Angeom}
\end{equation}
these are also the equations describing the moduli space of the $N=1$ theory in four dimensions for the gauge group $U(1)^n$ with the given superpotential. It is a Calabi-Yau geometry. In this case we have that $U=X^n$, $V=Y^n$, $Z$ is as above and $W$ is determined as in \cite{Berenstein:2002ge}.

Again, we see that the complex structure is determined entirely by integers and that it provides a set of examples of fixed complex structure by fluxes.

We are interested in understanding two questions. First, we want to ask if there are analogs of Seiberg dualities that preserve this geometry, and how do the $k_i$ change when we do that. We also want to ask if we can show that after the Seiberg duality is performed we can match the spectrum of monopole operators between the two theories.  Secondly, we want to ask what happens when all the $k_i$ are different from zero, but in the case where the geometry described by \eqref{eq:Angeom} is singular.
We expect on general grounds based on the intuition from flour dimensions that when there are singularities in this geometry that there should be fractional branes appearing and these can lead to new branches in moduli space with branes exploring the singularities. In these situations the variety describing the moduli space is reducible.

The simplest example of a reducible variety is given by the curve $xy=0$ in $\BC^2$. It has two components $x=0$ and $y=0$. These join at $x=y=0$. If we have a
supersymmetric theory with a chiral ring whose relations are that $xy=0$, these indicate that the OPE of $xy$ has no chiral ring element on the right hand side. We also want to understand relation like this one when $x,y$ are monopole operators.

First, let us consider the general case with or without singularities. We will first consider the duality on one node as described in \cite{Giveon:2008zn} for a theory with $\mathcal{N}=3$ SUSY, $U(N)$ gauge group and level $k$ and $N_f$ flavors. The main observation is that $N_f$ stays fixed,  $N\to N_f-N+|k|$ and that $k\to -k$.  The change of rank does not affect the general discussion of the algebra associated to a quiver theory. The change $k\to -k$ affects the superpotential so  in principle it can change the moduli space of vacua.
In the theories we are studying, we would like to make the substitution $k_i\to -k_i$ for some $i$. Obviously this will not preserve the condition $\sum k_i=0$. Also it seems natural not to touch the nodes $\alpha$ with $\alpha>i+1$, or $\alpha <i-1$, because they would leave the terms in the equation \eqref{eq:Angeom} invariant.

Our options are to change $k_{i-1}\to k_{i-1}'$ and $k_{i+1}\to k_{i+1}'$ while also changing $k_i \to -k_i$, so that the three numbers $k_{i-1}$, $k_{i-1}+k_i$ and $k_{i-1}+k_i+k_{i+1}$ appearing in
the right hand side of \eqref{eq:Angeom} are permuted into each other. It is easy to see that the solution is to send $k_{i\pm1}\to k_{i\pm1}+k_i$. Then only two of the monomials are exchanged. This is an exchange of two roots in the equation \eqref{eq:Angeom}. We will call this operation a Weyl reflection by analogy with how similar dualities work in four dimensions \cite{Cachazo:2001gh}.

Now, for these theories the description of the basic monopole operators are more complicated than those of the $A_2$ quiver. First, generically, we will have a common flux equal to either plus one or minus one on all the nodes for a single brane. Next, we need to consider how the simplest monopoles will look like. Again, we can use the coloring trick for the nodes and color them according to whether they are sinks or sources. From sources a net number of arrows will emanate. For sinks a net number of arrows will
fall in. We will require that there are no closed loops: these closed loops are multiples of either $U,V$ or linear combinations of $W,Z$ multiplying a simpler operator. Again, there are operations that let us build other basic monopoles: we can replace a path in one direction from node $i$ to node $j$, by a path from node $i$ to node $j$ in the opposite direction, and remove closed loops. The counting of these possibilities is hard to do.

Again, for the set of these operators, we can change the sign of the flux and reverse all the arrows and this way we get a new allowed monopole operator. Thus they always come in pairs, $P_{\alpha}, \tilde P_{\alpha}$ so similar relations will hold: $P_\alpha \tilde P_{\alpha} = h_\alpha(W,Z)$. It is also easy to see that if the $k_i$ are not coprime, but instead we have that $k_i = \ell k_i'$, then the resulting set of allowed operators gives an orbifold by $\BZ_\ell$ of the variety associated to  $V_{k_i}'$. Also, $P, \tilde P$ again give us a $\BC^*$ bundle over the three dimensional Calabi-Yau  base.

Let us now perform a Seiberg-like  duality as described above. We will show that for each $P_\alpha$ in the first theory we can find $Q_\alpha$ in the second theory with the same quantum numbers, and that it is such that $P_\alpha \tilde P_\alpha \simeq Q_\alpha \tilde Q_\alpha$, in that they give rise to the same polynomial $h_\alpha$. Assume this is at node $i$, and let us assume for simplicity that
\begin{equation}
P_\alpha = \dots y_i^{t_1} x^{t_2}_{i-1} \dots
\end{equation}
Then in the other theory with the same labeling of the nodes we take the wave function\footnote{After we Seiberg dualize the quiver we invert the arrows: $x_{i}'\sim x_{i}^{\dag}$ and so on. Here we rename the variables such that the tilde variables denote the bifudamental corresponding to the field going in the same direction as in the original quiver, i.e. $\tilde{x}_{i}\sim y_{i}'$, etc.}
\begin{equation}
Q_\alpha=  \dots \tilde x_i^{t_2} \tilde y^{t_1}_{i-1} \dots
\end{equation}
This is, we do it by exchanging
\begin{eqnarray}
(x_i, y_i) & \to& (\tilde x_{i-1}, \tilde y_{i-1})\\
(x_{i-1}, y_{i-1}) &\to& (\tilde x_{i}, \tilde y_{i})
\end{eqnarray}
This exchanges incoming arrows by outgoing arrows but it is clear that this is very different than how it happens in four dimensions, because the substitutions here involve an extra node in the quiver.

This procedure is depicted graphically in figure \ref{fig:ANflip}
\FIGURE{\includegraphics[width=7 cm]{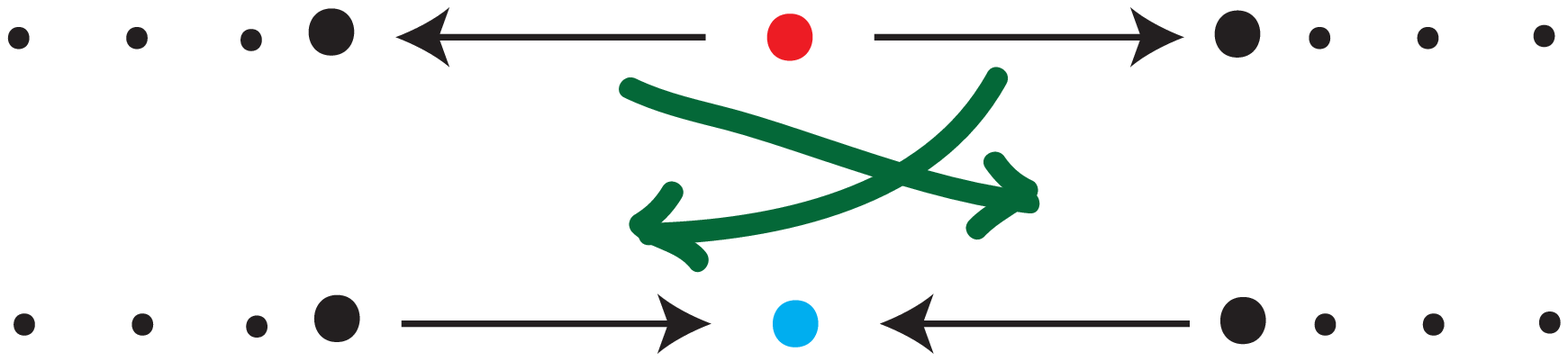}\caption{The procedure of building a new monopole wave function by flipping arrows after a duality move. }\label{fig:ANflip}}

 Consistency at node $i$ implies that $t_1+t_2+k_i=0$, and we see that when we switch $k_i\to -k_i$, the consistency condition at node $i$ is maintained.
It is also easy to show that if nodes $i-1$ and $i+1$ are consistent after doing the duality, so the $Q_\alpha$ is an allowed monopole wavefunction. Also, we should notice that in the equation defining \eqref{eq:Angeom} we have that $x_iy_i= (W+(k_1+\dots +k_i) Z)$, so that the roots in the polynomial $h_\alpha$ get permuted the same way that
they are permuted in the Calabi-Yau base.

It doesn't take much effort to convince oneself that $Q_\alpha \tilde Q_\beta$ also coincides with $P_\alpha \tilde P_\beta$ at the level of polynomials of $U,V, W, Z$.
Thus the varieties that define the non-singular bulk geometry are the same. Thus we could claim that the two theories are Seiberg dual at least at the level of sharing the moduli space geometry for branes in the bulk.

However, if the gauge groups $U(N_{i-1})$ and $U(N_{i+1})$ where to be thought of as global symmetries for the basic fields in the theory, the associated  flavor structure that sets up the isomorphism would not reverse the arrows on themselves, so the global symmetry content of the way we construct the operators has changed. This is different than what happens in four dimensional setups \cite{Seibergdual}.

We still need to consider what happens in the case where we have singularities in codimension $2$ in the geometry and what is the fate of these Seiberg-like dualities in these cases.

First, we can see that the equation \eqref{eq:Angeom} leads to a  singular variety with singularities only where $U=V=0$. Furthermore, there is always a singularity at $Z=W=0$. But there can be non-isolated singularities. If we fix $Z\neq 0$, these occur only if $\partial_W C=0$ at the same time that $C=0$ and these occur only if there is at least one repeated root. This implies that necessarily we have $k_1+\dots +k_{i-1} = k_1+\dots +k_j$. So that $k_{i} +\dots +k_j=0$. This is, we have a subset of
consecutive nodes such that their levels add to zero.

Two things happen then. First, there are fractional brane branches of the moduli space associated to reduced rank gauge groups in the associated four dimensional theory (these can be explicitly constructed from field theory arguments alone \cite{Berenstein:2002ge}). There are also new gauge invariant operators that are also not associated with elements of the center of the algebra \cite{Berenstein:2000ux} and can only be evaluated to be non-zero when the branes split at the singularity.

We can also check that there are fractional monopole operators. These are associated to some nodes having magnetic flux turned on and some off. For example, in the split above, where $k_i+\dots +k_j=0$, we can have the flux configuration depicted in figure \ref{fig:fluxconfig}.
\FIGURE{\includegraphics[width=7cm]{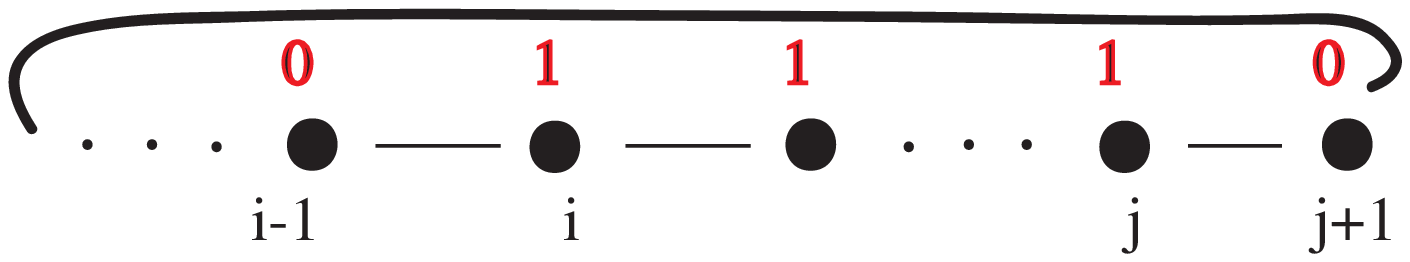}\caption{Flux configuration for fractional monopole setups. The flux values are shown in red. }\label{fig:fluxconfig}}
The figure shows that the fluxes get turned on exactly for the nodes that participate in the sum $k_{i} +\dots +k_j=0$.

Now, when we consider the associated minimal wave functions all the arrows connecting the nodes with flux to those with zero flux must be absent. Also, the operations that let us move a path in one direction from node $\alpha \to \beta$ to going around the quiver in the opposite direction are forbidden. Hence the minimal fractional monopoles are unique. Let us call these $F_i$ and $\tilde F_i$. It is easy to show that these have R-charge given by
\begin{equation}
[F_i] = (|k_i|+|k_i+k_{i+1}|+|k_{i}+k_{i+1}+k_{i+2}|+\dots +|-k_j|)/2
\end{equation}
(remember each $x_\alpha, y_\alpha$ have R-charge $1/2$) and it is also obvious that they are located at the singularity: when we multiply by U or V we find that arrows that would go forming a loop around the  quiver would need
to be turned on, but arrows connecting nodes with different fluxes are absent. Thus $UF_i= V F_i=0$. This is, it is localized where $U=V=0$ and they are naturally associated to the singularity as well. There are similarly some other monopole operators where the nodes that have flux in $F_i$ have zero flux, and the nodes that have zero flux have flux equal to one. These would be $F_{j+1}$ and $\tilde F_{j+1}$.

Also notice that $F_i \tilde F_i$ is a function only of $Z$, as when $U=V=0$ the only free coordinate is $Z$ (the variable $W$ is determined by the repeated roots).
Indeed, one can show that $F_i \tilde F_i = Z^{2 [F_i]}$ so the moduli space of the fractional branes is described by an ALE space ( an $A_s$ singularity), indeed $\mathbb{C}^{2}/\mathbb{Z}_{2 [F_i]}$, whose quotient is determined by the
Chern-Simons couplings. Because of $\mathcal{N}=3$ supersymmetry we know that the moduli space has to be hyperkhaler and it is of smaller dimension than the bulk (which has dimension 8). Indeed, this suffices to know the topology of the fractional brane branches of moduli space completely.

Now let us perform the Seiberg-like duality on node $i$. Clearly, now we have that $k'_{i+1}+\dots +k_j=0$, since $k_{i+1}'= k_i +k_{i+1}$. So the nodes that have flux turned
on change, and the new monopoles have dimensions
$[ F'_i]=(|k_{i}+k_{i+1}|+|k_{i}+k_{i+1}+k_{i+2}|+\dots +|-k_j|)/2\neq [F_i]$.
Notice also that the `flipping of arrows procedure' does indeed disconnect node $i$ from node $i+1$, but that it leaves something hanging between nodes $i-1$ and $i$ that would violate the flux equations for node $i$ and node $i-1$, which now has been set to zero.

We thus find that the topology of the fractional brane branches changes between one theory and the other and  that even the spectrum of dimensions of operators changes. Naively, one would assume that this implies that the duality fails. remember however that this is a classical result that might be subject to quantum corrections.
Indeed, we will see in the next subsection that this will be fixed by quantum corrections\footnote{We thank A.Kapustin, C. Closset and S. Cremonesi for pointing to us that these corrections were possible and for various enlightening discussions about this computation.}.


\subsection{Quantum corrections to monopole charges}\label{sec:quantcorrAn}

In this section we will show that the previously found fractional branches are actually coincident before and after duality, if we take on account quantum effects. We begin by recalling that monopole operators in the Euclidean theory in flat space are classical solutions of the form
\begin{equation}
F\sim *d\left(\frac{1}{|x|}\right)H
\end{equation}
where $H$ is an element of the Cartan subalgebra (CSA) of $Lie(G)$, where $G$ is the gauge group. For the case of the quiver gauge theories analyzed here $G=\prod_{i}U(N_{i})$ and so, the rank of the CSA is $\sum_{i}N_{i}$. We can write then $H=\{H_{i}\}$ with $H_{i}=(n_{1,i},\ldots, n_{N_{i},i})$, and $n_{\ell,i}\in \mathbb{Z}$ for all $i$ and for all $\ell$ \footnote{This is essentially the quantization of monopole charges for a gauge group $G$ \cite{Goddard:1976qe}. Here we are assuming that all the $n_{\ell,i}$'s are integers, which is true for generic brane charges $\{N_{i}\}$, however for special cases it may be possible for the $n_{\ell,i}$ to take rational values, as in \cite{Berenstein:2009ay}.
This case will not be addressed on this paper.}.The first index indicates the diagonal component of the $U(N_i)$ matrix that gets a vacuum expectation value (as in equation \ref{eq:fluxtwo})
and the second index indicates the gauge group. The monopole operator $\mathcal{O}_{H}$, as an element of the chiral ring correspond to a bare monopole $T_{H}$ plus matter fields, which we denote as $\phi_{H}$
\begin{equation}
\mathcal{O}_{H}=T_{H}\phi_{H}
\end{equation}
In this paper we have treated so far both $T_H$ and $\phi_H$ together as classical objects on the cylinder, and quantized this set of solutions. If one starts with $T_H$ classical and add matter as a quantum correction to satisfy Gauss' law, we get the structure above. This can obscure the role that the F-terms might play in considering elements of the chiral ring.

For a 3d $\mathcal{N}\geq 2$ CS theory with only bifundamental matter the R-charge of $T_{H}$ is given by \cite{Benini:2011cm}
\begin{equation}\label{quantmon}
R[T_{H}]=-\frac{1}{2}\sum_{X_{ij}}\left(R[X_{ij}]-1\right)\sum_{k=1}^{N_{i}}\sum_{l=1}^{N_{j}}|n_{i,k}-n_{j,l}|-\frac{1}{2}\sum_{i=1}^{N_{G}}\sum_{k=1}^{N_{i}}\sum_{l=1}^{N_{i}}|n_{i,k}-n_{i,l}|
\end{equation}
where, by $N_{G}$ we denote the number of vertices in the quiver. A bulk monopole or diagonal monopole, has the following flux configuration (denote $N=\min\{N_{j}\}$)
\begin{equation}
n_{i,l}=n_i \text{ \ \ for \ \ }i=1,\ldots,N\text{ \ \ and \ \ }n_{i,l}=0\text{ \ \ otherwise \ \ }
\end{equation}
a relatively simple computation shows that the R-charge correction of all these monopoles vanishes for the $\mathcal{N}=3$ quivers we are studying. We will illustrate this in the simplest example. For our purposes we are interested in the monopole of minimal charge, that is $n_{1,\ell}=\pm 1$ and zero otherwise, for all gauge groups (all will have the same sign, and the absolute values in the formula tell us that the result will be independent of this choice). Then, it is straightforward to show that for this particular configuration the only possible values of $|n_{i,k}-n_{i,l}|$ are $0,1$, and we just need to count them. We get that
\begin{eqnarray}
R[T_{H_{\mathrm{bulk}}}]&=& \sum_i  \frac{-1}2 \left(2\times \frac {-1}2 [(N_i -1)+(N_{i+1}-1)]\right) -\frac 12\left(2(N_i-1)\right) \\
&=&0
\end{eqnarray}
This tells us that the bulk moduli space computation we did seems to be completely free of quantum corrections and the classical result we found seems to be exact.
The cancellations involve the matter fields and the gauge groups against each other. They do not cancel automatically: the matter content is balanced just right with the gauge fields.

Now, consider again the fractional monopoles we found in the previous section, say $F_i$. We write $F_i=T_{i}\phi_{i}$. Recall the flux configuration of these monopoles
\begin{equation}
n_{s,l}=1\text{ \ \ for \ \ }s=1\text{ \ \ and for \ \ }l=i,\ldots, j\text{ \ \ and \ \ }n_{i,l}=0\text{ \ \ otherwise \ \ }
\end{equation}
The R-charge of the bare monopole $T_{i}$ is not zero, it's instead given by
\begin{equation}
R[T_{i}]=\frac{1}{2}\left(N_{i-1}+N_{j+1}-N_{i}-N_{j}\right)+1
\end{equation}
After Seiberg-like duality in the node $i$, the CS levels change as discussed previously and the rank at node $i$ as
\begin{equation}
N_{i}\rightarrow N_{i-1}+N_{i+1}-N_{i}+|k_{i}|
\end{equation}
Consider now the monopole denoted previously as $F'_i$, this has fluxes turned on, between nodes $i+1$ and $j$ only. A direct computation shows
\begin{equation}
R[T'_{i}]=R[T_{i}]+\frac{|k_{i}|}{2}
\end{equation}
the contributions to the R-charge, due to matter fields, were computed previously and after taking on account this correction it is clear that
\begin{equation}
R[F'_{i}]=R[F_{i}]
\end{equation}
matching the spectrum of monopole R-charges of fractional branches between the two dual theories.

Notice that this has an effect on the moduli space structure: the ALE spaces associated to fractional branes at the codimension two singularities necessarily depends on the ranks of the gauge groups. The classical result we found previously did not depend on these ranks. These are interpreted as quantum corrections to the topology of the moduli space. Instead of having an $A_s$ singularity as we computed before, we get an $A_{s+2 R[T_i]}$ singularity. This correction to the topology is obtained from matching R-charges in the product of the two operators $F_i$, $\tilde F_i$ with a power of $Z$ (we expect that the argument that  led us to $UF_i=VF_i=0$ still holds).
This is the only way to have the product of the two operators of positive and negative flux preserve the R-charge when we multiply them.

Now we find also that if we consider the two different fractional monopoles with positive flux $F_i,F_{j+1}$ , they have the same flux vector as a bulk monopole and the extra quantum correction to the  R-charge almost cancels, but not quite. There is a remnant equal to $2$. This indicates that a bulk brane does not automatically fractionate on reaching the singularity. There seems to be some binding energy that needs to be supplied to it to separate the two fractional branes. Since the binding energy is of order one, and the total energy is of order $k$ (the typical Chern-Simons coupling), in the large $k$ limit the binding energy is small and can be ignored on a first pass. It would be interesting to see how this affects the global structure of moduli space in further detail. This is beyond the scope of the present article.

\section{Reflection functors do not lead to Seiberg dualities}\label{sec:homol}

In the previous section we saw an example where the natural Weyl reflection preserved the moduli space. Here we want to check whether this is true in more general cases, or if the notion of Seiberg duality from four dimensions generally breaks down on going down to three dimensions in the presence of Chern-Simons terms.

Our purpose in this section is try to copy the natural notions of Seiberg duality in four dimensions so that it makes sense for three dimensional CS-matter theories, using as a guide the general form of Seiberg-duality \cite{Seibergdual}, developed in \cite{Berenstein:2002fi} and see if that works to generate a Seiberg dual theory. These dualities
contain the Weyl reflection dualities for $\mathcal{N}=2$ theories in four dimensions as special cases, which we analyzed in the case of the $\mathcal{N}=3$ theories in three dimensions.
Therefore these serve as a candidates to generate additional Seiberg-like dualities in three dimensions.  The advantage of this type of formulation is that it is systematic, and it does not depend on having $(p,q)$ five brane setups.   Also the set of ideas can be applied in more general contexts and is closely tied to topological field theory. An additional advantage is that  and one can also compute the  coupling constants from such a formulation.

The spirit of the paper \cite{Berenstein:2002fi} is that the operation of Seiberg duality can be described purely based in terms of homological algebra. We will call this the homological Seiberg duality. There, the reflection functor applied to a quiver category preserves the (bounded) derived category of modules of it. Let's review the basics of this duality. Each node of the quiver correspond to a bound state of branes (a bounded complex), call it $[B_{i}]$. Then in the simplest case, if there are $n_{i}$ arrows going from the node $[B]$ (the reflection node) to $[B_{i}]$, then the reflection functor is just a change of basis
\begin{eqnarray}\label{bdduality}
[B]&\rightarrow& [\overline{B}]\nonumber\\
{}[B_{i}]&\rightarrow& [\widehat{B_{i}}]\equiv [B_{i}]+n_{i}[B]
\end{eqnarray}
where $[\overline{B}]$ is the anti-brane of $[B]$. If there is adjoint matter one considers the maximal (left) extensions of $[B_{i}]$ by copies of $[B]$ and this leads to the
dualities first described by Kutasov \cite{Kutasov:1995ve}. Here the change of basis is of the type\footnote{Here, we are considering that the theory may have an adjoint field $\chi$ in the node $[B]$. In that case, for a superpotential of the form $Tr(\chi^{r+1})$ then $m_{i}=rn_{i}$ \cite{Berenstein:2002fi}}
\begin{eqnarray}\label{bdduality}
[B]&\rightarrow& [\overline{B}]\nonumber\\
{}[B_{i}]&\rightarrow& [\widehat{B_{i}}]\equiv [B_{i}]+m_{i}[B]
\end{eqnarray}
There is a similar version for incoming arrows instead of outgoing arrows. This corresponds to taking extensions on the right. Thus, we have to make a choice of whether we choose to make a reflection by maximal left extensions or maximal right extensions. Notice that only nodes with either incoming arrows to the reflection node (or outgoing arrows) are changed. Both reflections lead to the same dual quiver theory, but the brane content of the brane bound states is in general different between them.

 If the charges of the branes (the rank of the gauge groups) $([B],[B_{i}])$ are $(N,N_{i})$, then this leads to the change of ranks
\begin{eqnarray}
N&\rightarrow&\tilde N =\sum_{i}N_{i}m_{i}-N.
\end{eqnarray}
and all the arrows from the reflection node to $B_i$ are reversed.

This leads to two quiver theories with the same derived category of modules.
Two quivers whose derived category of modules is equivalent share the same center and the geometry associated to the center corresponds to branes in the bulk in the four dimensional theory. Thus the reflection operation guarantees that the moduli space of branes in the bulk matches in four dimensions. Also, the branches associated to a bulk brane splitting at singularities match, but not all branches can be mapped into each other. This can be seen from duality cascades where some branches with D-branes in the bulk plus fractional branes do not match under this operation, as the duality cascades reduce the general number of branes in the bulk. This is fine: Seiberg dualities do not have to match all branches of moduli space, only those that are connected to each other. Gaugino condensation can remove some of these connections, so one can have dualities that only map some branches of one theory to some branches of another theory. At the level of categories of modules it is also not true that every module in one algebra corresponds to a module in the derived equivalent algebra \footnote{Such an equivalence is a Morita equivalence, which is a stronger equivalence: that the category of modules matches}: it corresponds to an object in the derived category. This accounts for different stability properties of branes depending on FI terms \cite{Douglas:2000ah}.

A different aspect of this duality that is often overlooked is that the change of basis can also be used to compute the gauge coupling constants of the dual objects.
The duality states that
\begin{eqnarray}
\frac{1}{\tilde g^2} &=& -\frac 1{g^2}\\
\frac 1{\tilde g_i^2}&=& \frac 1{g_i^2} +\frac {m_i}{g^2}
\end{eqnarray}
and it usually happens when the RG flow of the theory force the gauge coupling constant of the node we are dualizing to become infinite (or the real part of the square inverse coupling constant to vanish). The analytic continuation to negative values is interpreted in terms of the brane becoming anti-BPS, so that replacing branes by anti-branes
leads to a BPS configuration where a supersymmetric theory describing the low energy dynamics is available. Notice that at the strong coupling singularities $1/g^2=0$
we have that $\tilde g_i = g_i$. A double reflection by first left extensions and then right extensions brings us back to the same theory.

 How we can try to possibly extend this duality to three dimensions?. In \cite{Aganagic:2009zk} it was shown that each node of a quiver gauge theory on the worldvolume of a 2-brane probing a CY four-fold $X_{4}$ singularity correspond to a vanishing cycle $[\Delta_{\alpha}]$ of the CY three-fold $X_{3}$ seeing it as the base of the complex line bundle $X_{4}\rightarrow X_{3}$. The CS levels corresponds to topological charges associated to each $[\Delta_{\alpha}]$ in the presence of non trivial RR-flux $F_{2}$. To be more precise, each cycle can be written as a sum of even homology classes of $X_{3}$ (see also \cite{Cachazo:2001sg}), say $[\Delta_{\alpha}]=Q^{6}_{a}[D_{a}]+Q^{4}_{i}[C_{i}]+Q^{2}[pt]$, where $Q^{p}_{a}$ is the $p$-brane charge. The integral of the RR-flux over the different sub-cycles coupled to the branes via a CS action gives the levels. Hence the change of basis (\ref{bdduality}) can be seen as a change on the basis of cycles and, therefore, in CS levels
\begin{eqnarray}\label{bddualitylevels}
k&\rightarrow& -k\nonumber\\
k_{i}&\rightarrow& k_{i}+m_{i}k.
\end{eqnarray}
which is the same operation on coupling constants in three dimension as the corresponding four dimensional dual.

Is interesting to note that the change of basis (\ref{bdduality}) can be performed alternatively using the arrows that have their head at $[B]$ instead of the tail. This will give the same dual quiver and superpotential as if we would use (\ref{bdduality}). The only difference may be in the change of the ranks, but taking on account the CS levels, if $n_{i}$ arrows go from $[B_{i}]$ to $[B]$
\begin{eqnarray}
k&\rightarrow& -k\nonumber\\
k_{i}&\rightarrow& k_{i}+\tilde m_{i}k.
\end{eqnarray}
which will give in general a different set of CS levels than using (\ref{bddualitylevels}). At this point we can only say that the resulting three dimensional theories will share the same $X_{3}$ base of the CY four-fold, but nothing more. Obviously,  it is trivial to check that the dual pairs proposed in \cite{Giveon:2008zn} and \cite{Amariti:2009rb} are included in our prescription. However in those cases, the existence of a known brane construction for the theory allows to show explicitly that the Hanany-Witten effect \cite{Hanany:1996ie} must be taken into account in the change of rank of the gauge group. This effect cannot be reproduced by our construction and it should be added by hand afterwards if one can show consistency of the theory when we look at monopole operators and the construction of the associated Calabi-Yau fourfold.

Let us consider now a single theory that corresponds to a chiral quiver (one where the net number of arrows from node $i$ to node $j$ is different from the number of arrows from node $j$ to node $i$), and let us also consider that the theory makes sense as a four dimensional theory that is anomaly free, etc. This implies that the number of incoming arrows is equal to the number of outgoing arrows. Now let us perform two consecutive dualities by left extensions for a particular node. This takes us from a quiver to itself, but at the same time we can check that the levels have changed as follows:
\begin{eqnarray}
k\to& -k &\to k\\
k_{i} \to & k_i +m_i k & \to k_i+m_ik- \tilde m_i k
\end{eqnarray}
 and this generally lets us move the Chern-Simons levels from one node to other nodes: the only case where this can not be done is in non-chiral quivers.

 Clearly we can repeat this operation an unlimited number of times, taking the Chern-Simons levels of some nodes to infinity in the process. In such theories we generally expect that the monopole operators will decouple (become infinitely heavy) and that the moduli space of the bulk therefore changes.
 One might still be able to save the day if by chance the  dimension of the fields is such that the fields go to zero dimension as the Chern-Simons level grows, but the general arguments we have given before suggest that the moduli space has changed in these operations as the charges of the monopole operators are different. Let us discuss this with an example.

Let us look at the one quiver theory we have analyzed already: the $\BC^3/\BZ_3$ quiver. In that theory after we perform these operations we find that the number of basic monopoles with flux one changes between one theory and its purported set of duals. The relations will also dictate that the dual pairs will have different chiral rings, as dictated by (\ref{eq:Z3rels}). Indeed, this should serve as a template for general chiral theories. We expect that the relations will also look generally like
  $F\tilde F \simeq C^{\sum a_i|k|}$ between fundamental monopole objects. The main reason for this type of relation is that the flux equations require more arrows to end at nodes with large Chern-Simons levels. Thus when we take a monopole with flux $1$ on each node times an (anti-) monopole with flux $-1$ we get in general of  order
  $|k|$ loops in the quiver. The mesonic branches are already mapped to each other by how the duality operates in four dimensions.
  This renders the set of relations completely different between candidate duals.

  Our main conclusion from this section is that the techniques that work in four dimensions to generate Seiberg-like dualities seem to break down completely in the three dimensional case, and there are perhaps some isolated examples where this does not happen. This does not preclude the possibility of Seiberg dualities in general. It just tells us that the techniques used to generate them don't work. We should be open minded about alternatives. Indeed, similar observations can also be found in \cite{Kapustin:2010mh} where naive Seiberg duality operations fail and it was suggested that they get fixed by hand. Further analysis has been done in \cite{Hwang:2011qt}. After this paper was released, a further work came out \cite{Benini:2011mf} that suggests a different recipe for changing the Chern-Simons levels between some of these theories to realize dualities.

  We will explore in the next section whether it is possible to have Seiberg-like dualities built by hand to match between two sets of four dimensional duals, but without requiring that the Chern Simons levels are matched by the brane change of basis.

\section{A Seiberg duality by hand}\label{sec:hand}

Here we will consider again doing a four dimensional Seiberg duality on the $\BC^3/\BZ_3$ quiver, but we will impose no restriction on the Chern Simons levels of the dual theory. Instead, we will  require that in the new theory we can find a one to one map between elements of the chiral ring as represented by monopole operators and the same class of monopole operators in the original theory. These can be computed given arbitrary Chern-Simons levels for both theories. We will then ask what implication this identification has for matching the Chern Simons coupling constants across duals. Our purpose here is to determine if there is a way out of the impasse we hit in the previous section, with the understanding that the matching is essentially by an ad-hoc procedure. One important point to remark is that due to the fact we are considering the theory with equal ranks for all gauge groups, the BPS monopoles do not suffer from quantum corrections according to equation (\ref{quantmon}) in the toric phase analyzed in section \ref{sec:c3z3}. Indeed the R-charge of the simplest diagonal bare monopole is given by (using the notation of section \ref{sec:c3z3})
\begin{equation}
-3(R[X_1]+R[X_2]+R[X_3]-3)(N-1)-3(N-1)=0
\end{equation}
where we used the fact that $R[X_1]+R[X_2]+R[X_3]=2$. When we consider the Seiberg dual quiver, the phase is no longer toric and in order to preserve spherical symmetry of the solutions we are required to turn on fluxes which will give us bulk monopoles that are not diagonal in the sense of \cite{Benini:2011cm}, for example\footnote{This is essentially because of the fact that F-term equations will force us to have solutions, equivalent to simple modules which have non-zero vevs in the off diagonal. Consider for example a bifundamental transforming in the $(\mathbf{2},\mathbf{1})$ representation, explicitly: $A=(A_{1},A_{2})$. Denote the fluxes in the gauge groups by $\mathrm{diag}(n_{1,1},n_{1,2})$ and $n_{2}$. Then equation (\ref{sphersym}) implies $n_{1,1}=n_{1,2}=n_{2}$ if both $A_{i}$ are generic, giving a non-diagonal flux configuration if $n_{2}\neq 0$.}. In the following we will consider the dual quiver with ranks $(N,N,2N)$ for $N=1$ only, for simplicity, however it is important to point out that the quantum corrections for the bare monopole R-charge vanish for $N>1$, indeed (for the notation, see below)
\begin{equation}
R[T_{H_{\mathrm{diag}}}]=-6(R[A]+R[B]+R[C]-3)(N-1)-\frac{1}{2}\left(2\times 2 (N-1)+8(N-1)\right)=0
\end{equation}  

The standard Seiberg dual of the four dimensional $\BC^3/\BZ_3$ theory is depicted in figure \ref{fig:C3-Z3sd}.
\FIGURE{\includegraphics[width=5cm]{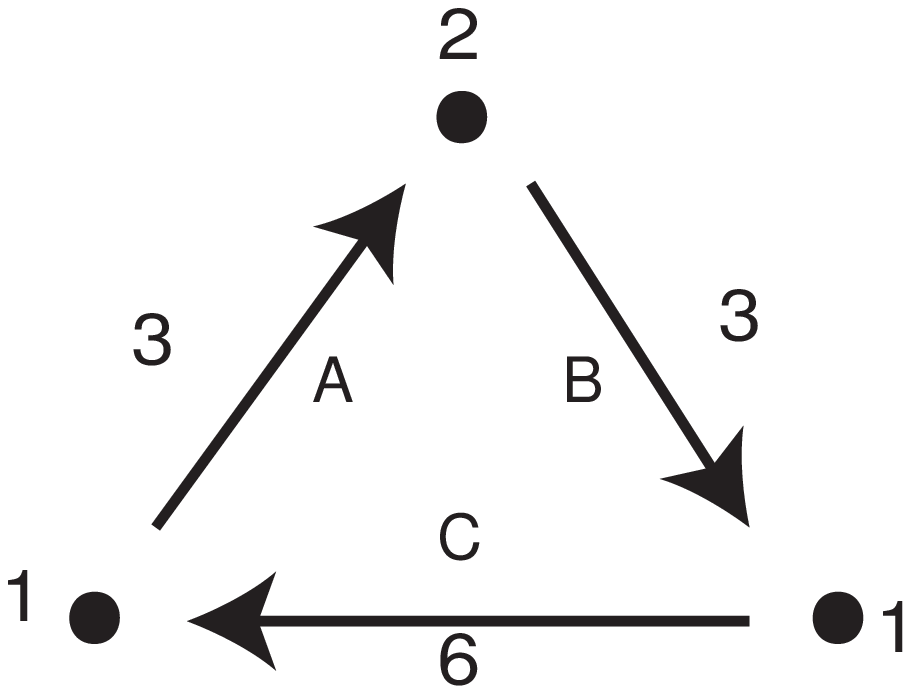}\caption{The Seiberg dual quiver to the four dimensional $\BC^3/\BZ_3$ quiver.}\label{fig:C3-Z3sd}}
The theory has three types of fields, which we have called $A,B,C$. The fields $A,B$ transform in the $\mathbf{\bar 3}$ of $SU(3)$ of the global symmetry and the field $C$ transforms in the $\mathbf{6}$ of $SU(3)$.

The superpotential is the unique single trace $SU(3)$  singlet that is contained in the $\mathbf{\bar3}\times \mathbf{\bar3}\times \mathbf{6}$ given schematically by
$\tr(CBA)$. The number on the nodes indicate the ranks associated to a brane in the bulk. Thus the fundamental representations are not products of $U(1)^3$.
The condition that is equivalent to the sum of the levels cancelling is given by $k_B+2k_C+k_A=0$ \cite{Berenstein:2009ay} where we label the nodes by the field in the opposite side of the triangle.

As shown in \cite{Berenstein:2002ge}, when considering elements of the center of the algebra, it matches with the original $\BC^3/\BZ_3$ quiver. This is a general property of quiver diagrams related by the reflection functors. These are schematically of the form $ACB+CBA+BAC$ and these completely characterize the representation theory of the algebra. We can take now these functions that are in the $\mathbf{10}$ of $SU(3)$ (they have a single row of three totally symmetric boxes in the Young tableaux of $SU(3)$ representations)  and show explicitly that their powers have representations consisting of $3 k$ boxes in the totally symmetric representation of $SU(3)$. Now we want to understand the spectrum of monopoles in this situation. Notice that the factors of $2$ in the rank make our monopole equations more complicated. If $A$ is in the fundamental of $U(2)$, labeled by $A^i$, $i=1,2$, and $B$ is in the antifundamental $B_i$, $i=1,2$, then the flux equations are given by a $4\times 4$ matrix
\begin{equation}
F\begin{pmatrix} k_A&0&0&0\\
0&k_B&0&0\\
0&0&k_C&0\\
0&0&0&k_C
\end{pmatrix}\psi = \begin{pmatrix}  B\partial_B-C\partial_C&0&0&0\\
0&C\partial_C-A\partial_A&0&0\\
0&0&A^{1}\partial_{A_1}-B_1\partial_{B_1}&A^{1}\partial_{A_2}-B_2\partial_{B_1}\\
0&0&A^{2}\partial_{A_1}-B_1\partial_{B_2}&A^2\partial_{ A_2}-B_2\partial_{B_2}
\end{pmatrix}\psi
\end{equation}
We can clearly see that the generators of the $SU(2)$ gauge symmetry inside $U(2)$ annihilate $\psi$, these are given by
\begin{eqnarray}
L_3 &\simeq& A^{1}\partial_{A_1}-B_1\partial_{B_1}-(A^2\partial_{ A_2}-B_2\partial_{B_2})\\
L_+&\simeq& A^{1}\partial_{A_2}-B_2\partial_{B_1}\\
L_-&\simeq& A^{2}\partial_{A_1}-B_1\partial_{B_2}
\end{eqnarray}
Here remember that $A,B,C $ also transform under the $SU(3)$ global symmetry and that this is implicit.
Thus $\psi$ is $SU(2)$ invariant, but the left hand side shows that it carries $U(1)$ charge under the diagonal $U(1)\subset U(2)$. Indeed, it carries charge $2k_C$ under the normalization where the $A$ carries charge $1$ and the $B$ carries charge $-1$.

Again a minimal monopole will depend on the details of the $k_{A,B,C}$. We have two options to consider, where $k_{A,B}$ have the same sign, or opposite signs, subject to the constraint that $k_A+k_B+2k_C=0$ and that none of them is zero. Consider first the case where the signs of $k_{A,B}$ are opposite and let us assume that $k_A$ and $k_C$ have the same sign and are positive (the choice of overall sign for the $k$ can be changed if we change $F\to -F$, so it can be chosen by hand). Then the minimal monopole with flux $1$ will have nodes $C$ and $A$ act as sources and node $C$ as a sink. To satisfy the charge constraints we find that the antimonopole wavefunction must be schematically of the form \begin{equation}
\tilde{M}\simeq B^{2k_C} C^{2k_C+k_A}.\label{eq:sdbasicmono}\end{equation}
 But remember that the $B$ carry $SU(2)$ charge, so the monopole operator must be $SU(2)$ invariant.
This requires that the group index structure of $SU(2)$ is handled correctly. The minimal $SU(2)$ invariant is of the form
\begin{equation}
B^a_1B_2^b-B_2^aB_2^b \simeq (B^a\wedge B^b)
\end{equation}
where
$a,b$ are flavor indices. Notice that this is antisymmetric in $a,b$. Since these $a,b$ indices are in the $\mathbf{\bar 3}$ of $SU(3)$, we find that these objects transform in the $\mathbf{3}$ of the $SU(3)$ flavor symmetry, so each of them carry a single flavor box of the $SU(3)$ symmetry . Similarly, $C$ carries two boxes. In total between the $C$ and
the $B\wedge B$ combination there are $2 (2k_C+k_A)+ k_C$ boxes. Also notice that since there are products of $B$ and $C$ fields, the flavor information can be moved between them by using the F-term relations for $A$. One can show in simple examples (for example $k_C=k_A=1$) that this implies that the flavor structure is completely symmetric. We will now assume that this is always true: that only the completely symmetric representation with $5k_C+2k_A$ boxes appear. This should be good also for the case where the flux is not just the minimal flux, but for arbitrarily large flux.

The basic monopole wave functions will have the shape
\begin{equation}
 M\simeq A^{2k_C+k_A}B^{k_A}\label{eq:sdbasicmonod}
\end{equation}
Again, $SU(2)$ gauge invariance requires that we take singlets. These are of the form $AB$ and $A\wedge A$, where $AB$ uses the standard matrix contraction. The combinations $A\wedge A B\wedge B$ can be written in terms of $AB$ contractions, because of the identity $\epsilon ^{ab} \epsilon_{cd} = \delta^a_c\delta^b_d-(a\leftrightarrow b)$. This avoids overcounting. Again, the $AB$ count as a single box of
$SU(3)$, as the F-terms relation for $C$ imply that combinations involving the $\mathbf{\bar 6}$ of $SU(3)$
are absent, and similarly we have that $A\wedge A$ counts as a single box of $SU(3)$. Thus we obtain a monopole with $k_C+k_A$ symmetric boxes. Also, one shows that
the product of $M\tilde M$ must consist of terms of the form $(ABC)^{k_A+2k_C}$ with all $SU(3)$ boxes symmetrized. The structure starts looking similar to
the one we described in the $\BC^3/\BZ_3$ quiver, in particular equation \eqref{eq:Z3rels}.

Now let us try to match this to the field theory described in the $\BC^3/\BZ_3$ by checking that the basic monopoles transform in the same way under the $SU(3)$ flavor structure. We must therefore match the flavor content of equations \eqref{eq:sdbasicmono} and \eqref{eq:sdbasicmonod} to \eqref{eq:MonoC3} and the relations \eqref{eq:Z3rels}. We get that we need to have
\begin{eqnarray}
k_A+k_C &=& 2k_2+k_1\\
k_A+2k_C &=& k_1+k_2\\
5k_C+2k_A&=&2k_1+k_2= 3(k_A+2k_C)-(k_A+k_C )
\end{eqnarray}
These are easy to solve. We get that $k_C=-k_2$ and that $k_A=3 k_2 +k_1$. However, remember that $k_2, k_1>0$ and that $k_A,k_C$ have the same sign. This is a contradiction. Thus we do not find a match. There is no candidate dual. These theories represent new theories that can not be realized in the $\BC^3/\BZ_3$ quiver.

On the other hand, we can consider the option where $k_A,k_B$ have the same sign, and then $k_C$ has the opposite sign, let us assume that it is positive. The basic monopoles will then be of the form
\begin{eqnarray}
M&\simeq& (B\wedge B)^{k_C} C^{-k_B} \\
\tilde M &\simeq & (A\wedge A)^{k_C} C^{-k_A}
\end{eqnarray}
where $M$  will have $k_C-2 k_B$ symmetric boxes and $\tilde M$ will have $k_C-2 k_A$ symmetric boxes. Similarly $M\tilde M\simeq (ACB)^{2k_C}$.
Again, we can try to match relations to find that we need to have that
\begin{eqnarray}
k_C-2 k_B &=& 2k_1+k_2\\
k_C-2 k_A&=&2k_2+k_1\\
2k_C&=& k_1+k_2
\end{eqnarray}
Now, we can solve for $k_{1,2}$ to find that
\begin{eqnarray}
k_1 = -2k_B-k_C\\
k_2=-2k_A-k_C
\end{eqnarray}
so that a duality might exist only when $k_C\geq |k_B|/2$ and $k_C\geq |k_A|/2$ simultaneously.

Also notice that in the other direction we need that
\begin{eqnarray}
k_B&=& \frac{-3k_2-k_1}{4}\\
k_A&=&\frac{-3k_1-k_2}{4}
\end{eqnarray}
so the levels $k_{1,2}$ need to satisfy some special division properties.

\begin{figure}[ht]
\begin{center}
\includegraphics[width=7cm]{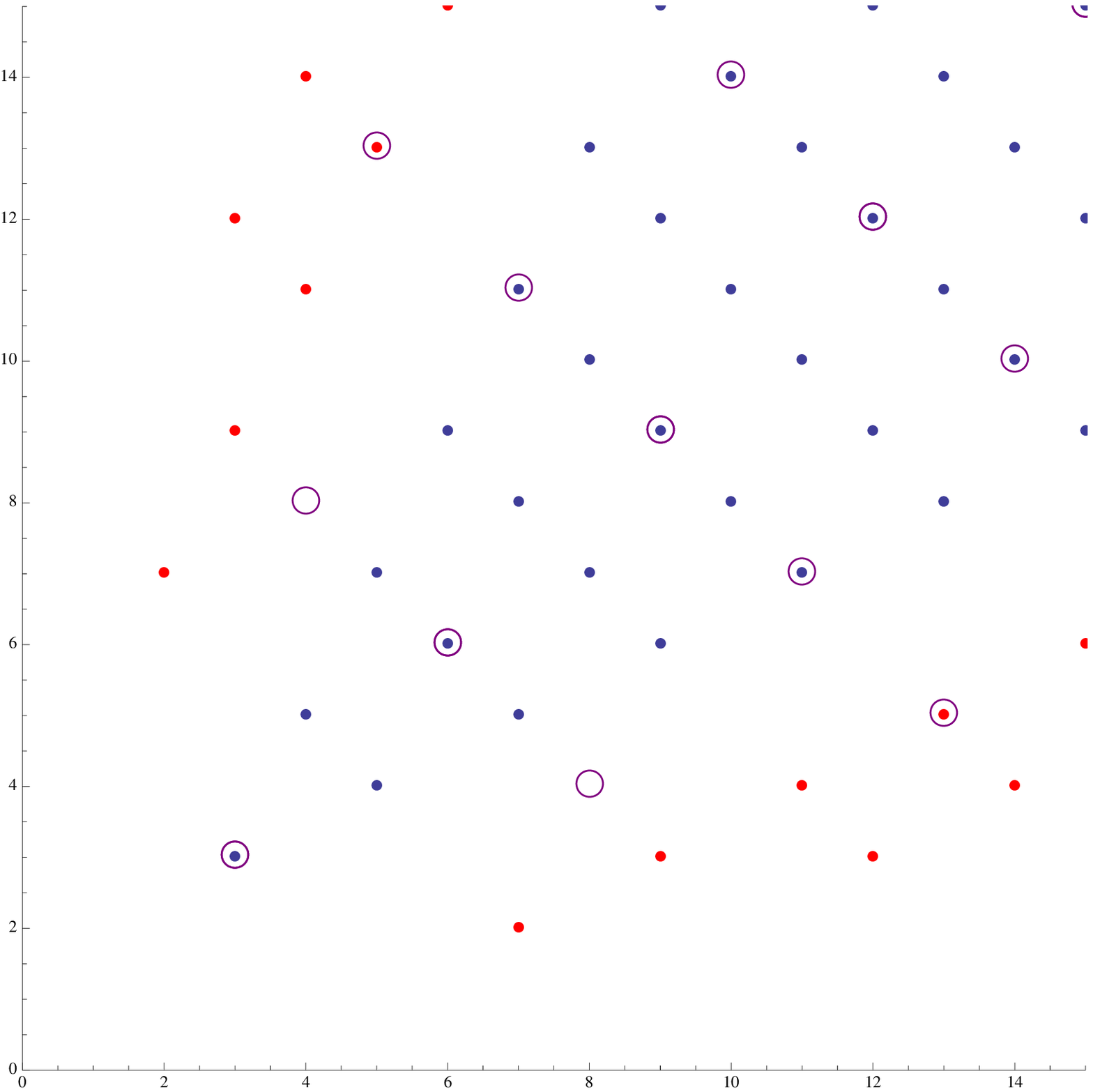}\caption{Covering of the monopole charges plane. The two coordinates indicate the number of $SU(3)$ symmetrized boxes of the two fundamental monopole operators $M, \tilde M$.
The blue dots indicate the theory with $k_1,k_2$.
The red dots indicate the setup with $k_A,k_B$ of opposite signs, and the purple circles indicate when $k_A,k_B$ have the same signs.}\label{fig:cover}
\end{center}
\end{figure}

It is interesting to plot the possible combinations of the $SU(3)$ color boxes for the monopoles in pairs $(Q[M],Q[\tilde M])$ where $Q$ counts the number of fundamental $SU(3)$ boxes in each of the minimal monopoles. This is in order to see which
combinations are realized giving a possible set of Seiberg dual pairs.

We find the result depicted in figure \ref{fig:cover}. First of all we find it curious that the theory
where $k_A,k_B$ have the same signs\footnote{For $k_A,k_B<0$, $(Q[M],Q[\tilde M])$ corresponds to the points $\{n_{1}(1,5)+n_{2}(5,1)+(3,3)+\varepsilon(3,3)| n_{i}\in\mathbb{N}, \varepsilon\in\{0,1\}\}$.} overlaps with both the regions covered by those where $k_A,k_B$ have opposite signs\footnote{For $k_{A},k_{C}>0$, $(Q[M],Q[\tilde M])$ corresponds to the points $\{n_{1}(1,2)+n_{2}(1,5)+(2,7)| n_{i}\in\mathbb{N}\} $ or if $k_B,k_C>0$ to the mirror under the diagonal of thee configurations.} as well as the theories defined by the original $\BC^3/\BZ_3$ quiver, defined by $k_1,k_2>0$\footnote{For $k_1,k_2>0$, $(Q[M],Q[\tilde M])$ corresponds to the points $\{n_{1}(1,2)+n_{2}(2,1)+(3,3)| n_{i}\in\mathbb{N}\}$.}. The fact that red and blue do not overlap is our observation that there is no possible duality between them, as we showed previously by contradiction. Moreover, the theories with $k_A,k_B$ of the same sign cover both regions but only on a sub-lattice: not all theories are covered even if they are defined on similarly allowed regions.

We find that the possible existence of Seiberg-like dualities behaves rather strangely: first, the map of levels from one theory to a possible dual can involve fractions and
there are new theories that are discovered this way that where not part of those that arise from the original four dimensional quiver of the $\BC^3/\BZ_3$ singularity with arbitrary $k_1, k_2$. We can also speculate that if we continue thinking about these theories as is done with duality trees \cite{Franco:2003ja}, we might be able to discover new theories as we keep on performing dualities and that these theories cover the quadrant of positive charges given above. This is interesting to study further.

\section{Hilbert series and the R-charges of the chiral ring operators}\label{sec:Hilbert}

So far we have assumed that the R-charges of the fields are known in advance and we have used this information implicitly to make calculations of the chiral ring.
However, in practice, most interesting interacting conformal field theories are not in the free field regime and the R-charges of the fundamental fields are unknown except perhaps as related to each other by the symmetries of the lagrangian, both discrete and continuous. In the example of the ${\cal N}=3 $ theories, the R-charge is $SO(3)$.
Hypermultiplets have two component chiral superfields $X, Y$ of opposite charges and they are related to each other by stating that $X$ and $Y^\dagger$ are related to each other by being a doublet under the $SO(3)\simeq SU(2)$ global R-charge symmetry. Thus the R-charge of $X$ is automatically determined to be equal to $1/2$. Similarly, the vector multiplet has three scalar fields that are in a triplet of $SO(3)$. The chiral partner of the ${\cal N}=2$ vector superfield, which we have called $Z$ in previous sections, naturally has R-charge equal to one. This is also determined from the superpotential. However, let us assume that we are not told that $X,Y$ have the same R-charge. Is there a way to determine the R-charges from just the ${\cal N}=2$ theory? In four dimensions, there is a process of a-maximization \cite{InW} that uses the anomaly properties of a hypothetical R-charge to compute the correct R-charge assignments.

The R-charge is constrained so that all terms in the superpotential of a conformal field theory are marginal. We require the same condition here. However, this does
not automatically tell us that the R-charge of $X$ and $Y$ are the same. Indeed, since $X,Y$ are not gauge singlets, given an R-charge assignment, we can consider a modified R-charge of the
form $\tilde Q_R = Q_R + \sum \alpha_i Q_i$, where $Q_i$ are the charges of the fields $X,Y$ under the diagonal $U(1)$  of the $U(N)$ nodes. In four dimensions this does nothing, since gauge invariance would require that for every operator the net charge cancels.

If it wasn't for the Chern-Simons modification of the Gauss law constraint, this would not affect the R-charge of any of the operators at all in three dimensions either. Indeed, we find that since the sum of the $Q_i$
at each node is $k_i$ times the flux at the node, the net modification of the charge of the given operator is proportional to the net common flux of a monopole operator. This is, our true freedom is reduced to counting how much charge a monopole operator carries for example. This is because only gauge invariant operators count for defining the theory.

Indeed, in the $A_{n-1}$ quiver example, we can use this freedom to choose an R-charge so that there is a unique common R-charge for all $X$, $R_X$, and a unique common R-charge for all $Y$, $R_Y$ constrained so that $R_X+R_Y=1$. This constraint is from the condition that the superpotential is marginal.
Notice that via this condition the $R$ charge of operators like $\tr(U) = \tr(X^n)$ would be $nR_X$ and the one for $V$ would be $nR_Y$, but it does not follow that $R_X=R_Y$ without using the knowledge of the $SU(2)$ R- symmetry of the lagrangian. However, the charge of $\tr(U)$ and $\tr(V)$ are constrained so that their product has charge exactly equal to $n$. This is, we are also constrained by the chiral ring relations. They must be consistent so that the products of $U$ and $V$ match in dimension to other elements of the chiral ring that we actually know about.

However, given our basic monopole operators, we can count the number of $X$ and $Y$ fields and give the operator the classical charge $n_xR_X+n_yR_Y$. There is a possibility of some additional quantum contribution, but this is not part of the classical analysis we have done so far (for recent work on this direction see \cite{Benini:2011cm}). Understanding if such a modification arises is beyond the scope of the present paper, since we are dealing mostly with classical solutions to the equations of motion: the vevs are such that the classical analysis is reliable. This requires that there are massless particles that feel the magnetic monopole background. The classical vevs in our cases of interest are supposed to Higgs the theory leaving no such massless particles behind.

Since there is no anomaly condition, we have to look elsewhere for a way to compute these R-charges. Ideally, we would do this using only the information of the chiral ring.
indeed, there is an alternative to understanding the a-maximization procedure, which is looking in dual supergravity backgrounds at Einsteins equations in the presence of supersymmetry. This gives rise to the volume minimization condition to compute the R-charge. However, we run into the problem that we do not have access to this geometry directly: we have the cone over the geometry as described by the moduli space of a brane in the bulk and we have the holomorphic functions on this cone.
The holomorphic functions on the cone is exactly the chiral ring we have been computing,

The one thing we can do is count the number of holomorphic functions of R-charge equal to $r$, let us call it $h_r$ and put them on a generating series.
The series we get is given by
\begin{equation}
H(t)= \sum_r h_r t^{2(r)}
\end{equation}
This series is the Hilbert series for the geometry. Since the $r$ charge is unknown, except for the constraints that we have, we find that $H$ actually depends on the choice of R-charge assignments. Also notice that $H(0)=1$ (there is an identity function corresponding to the vacuum) and also we find that $H(1)=  \sum_r h_r =\infty$.
$H$ is actually a meromorphic function at $t=1$. The order of the pole encodes the dimension of the moduli space.

What is most interesting is the residue of the pole of maximum order at $t=1$. This is proportional to the volume of the base of the manifold, after all the manifold is a cone. We have chosen the normalization so that if a free field has $R$ charge $1/2$, then the power attached to it is equal to $1$. This way the Hilbert series of $\BC^4$ is equal to $(1-t)^{-4}$ if each of the generators has R-charge $1/2$.
The volume of the 7-sphere is given by $\pi^4/3$, so we have that in general the proportionality constant of the leading pole behavior of the pole is given by
 \begin{equation}
H(t)= \sum_r h_r t^{(2 r)}\to \frac{ 3 vol}{\pi^4} (1-t)^{-4}+ O((1-t)^{-3})
\end{equation}
now, remember that the volume depends on our assignment of $R$ charge. So it is possible to minimize $vol$ with regards to these choices and implement a volume minimization procedure. Another way to write the series is to consider
\begin{equation}
\tilde H(q)= \sum_r h_r q^{(r)}
\end{equation}
which for $\BC^4$ gives $ (1-q^{1/2})^{-4}$ instead. The expansion around the pole at $q=1$ in this form gives us
\begin{equation}
\tilde H(q) \simeq \frac {16}{(1-q)^4} +\dots
\end{equation}
so the volume is given by the residue of the pole as follows
\begin{equation}
vol = \frac{\pi^4}{3\times 16} \hbox{Max. Res} (\tilde H(q) )|_{q=1}
\end{equation}
We will use the generating series in terms of the $q=t^2$ variable in this paper.
In any case, we can compare the normalized volume to that of the seven sphere:
\begin{equation}
\frac{vol}{vol S^7} =  \frac{\hbox{Max. Res} H(q)|_{q=1}}{16}
\end{equation}

Let us now take the example of the $\BC^3/\BZ_3$ quiver, where we can count operators. Indeed, we had argued that in all those cases we would obtain
monopoles $S=M_+,T=M_-$, with $k_1+2k_2$ fields and $k_2+2k_1$ fields respectively and the meson fields $U$, which have R-charge $2$ and correspond to a weight equal to $4$. Here we are using the $SU(3)$ symmetry to declare that all fields belonging to the same $SU(3)$ representation have the same R-charge.

The general meson operator of the schematic form $U^k$ has $R$-charge $2 k$. So we have that the R-charge of $M_+$ and $M_-$ add up to $2(k_1+k_2)$.
If we call the R-charge of $M_+$ as $R_+$ and the $R$ charge of $M_-$ as $R_-$, we have that $R_++R_-= 2(k_1+k_2)$. The general element of the chiral ring
starts as
\begin{equation}
U^{m}M_{+}^{m_{+}}M_{-}^{m_{-}}
\end{equation}
but the relations $M_+M_-\propto U^{k_1+k_2}$ tell us that we can choose either $m_+=0$ or $m_-=0$.
For each fixed $m_+$, we have that the allowed functions  transforms in the totally symmetric representation of $SU(3)$ with
$s=3m+(2k_{2}+k_{1})m_{+}+(2k_{1}+k_{2})m_{-}$ boxes. These have $(s+1)(s+2)/2$ elements. Summing over $m+\geq 0$ and $m_-\geq 0$, we obtain that
\begin{eqnarray}
H_+(t)&=&\sum_{m,m_{+}=0}\frac{(3m+(2k_{2}+k_{1})m_{+}+1)(3m+(2k_{2}+k_{1})m_{+}+2)}{2}q^{(m_{+}R_{+}+2m)}\nonumber\\
&=&\Big[(k_{1}+2k_{2}+1)q^{R_{+}+4}((k_{1}+2k_{2}+2)(q^{R_{+}}+1)-6)\nonumber\\
&&-2q^{2+R_{+}}((k_{1}+2k_{2})^{2}(1+q^{R_{+}})+7(2-q^{R_{+}}))\nonumber\\
&&+(k_{1}-1+2k_{2})q^{R_{+}}((k_{1}-2+2k_{2})(q^{R_{+}}+1)+6)\nonumber\\
&&+2(1+7q^{2}+q^{4})\Big]/(2(q-1)^{3}(q+1)^{3}(q^{R_{+}}-1)^{3})\nonumber\\
\end{eqnarray}
where we have used the representation $t= q^{1/2}$.
A pole at $t=1$ is also a pole at $q=1$ of the same order.
The sum over $m_-$ is similar
\begin{eqnarray}
H_-(t)&=&\sum_{m,m_{-}=0}\frac{(3m+(k_{2}+2k_{1})m_{-}+1)(3m+(k_{2}+2k_{1})m_{-}+2)}{2}q^{m_{-}R_{-}+2m}\nonumber\\
&=&\Big[(k_{2}+2k_{1}+1)q^{R_{-}+4}((k_{2}+2k_{1}+2)(q^{R_{-}}+1)-6)\nonumber\\&& -2q^{2+R_{-}}((k_{2}+2k_{1})^{2}(1+q^{R_{-}})+7(2-q^{R_{-}}))\nonumber\\
&&+ (k_{2}-1+2k_{1})q^{R_{-}}((k_{2}-2+2k_{1})(q^{R_{-}}+1)+6)\nonumber\\
&&+2(1+7q^{2}+q^{4})\Big]/(2(q-1)^{3}(q+1)^{3}(q^{R_{-}}-1)^{3})\nonumber\\
\end{eqnarray}
The reader may be worried that we are overcounting operators. In fact we are, but the sum over the states with $m_{\pm}=0$ will have at most a pole of order $3$ in $t$, and therefore will be irrelevant for our computations. The reason to present it this way is because the expressions of the sums in the strictly correct range are more convoluted. However, in the case $k_{1}=k_{2}=1$ it can be checked to coincide with the particular case of $M^{1,1,1}$ (see for example \cite{Davey:2009qx}). Is clear from these expressions and the linear relation between $R_{\pm}$ that only the R-charges of gauge invariant operators can be obtained from the volume minimization procedure. The order $1/(1-q)^{4}$ term in the Hilbert series is given by
\begin{eqnarray}
-3(k_2+k_1)&\Big[&\frac{16}{3}k_{2}^{4}+(-4R_{+}+16k_{1})k_{2}^{3}+(\frac{52}{3}k_{1}^{2}+R_{+}^{2}-6R_{+}k_{1})k_{2}^{2}\nonumber\\
&+&k_{1}(8k_{1}^{2}-2R_{+}k_{1}+R_{+}^{2})k_{2}+\frac{4}{3}k_{1}^{4}+k_{1}^{2}R_{+}^{2})\Big]/(-2k_{1}-2k_{2}+R_{+})^{3}R_{+}^{3}\nonumber\\
\label{eq:polyvol}
\end{eqnarray}
where we have used the relations between $R_+$ and $R_-$ to leave it in terms of the only variable: the R-charge of the minimal monopole, $R_+$.
Then, the values of $R_{+}$, are given by the zeroes of a cubic polynomial. Only one of them is real. It seems that only for the special combination $k_{1}=k_{2}$, the R-charge of $M_{+}$ is rational i.e. the Sasaki-Einstein manifold is regular. The value of $R_{+}$ has  a nice linear growth in $k_{1,2}$, as we illustrate for some values on the table below
\begin{center}
\begin{table}
\begin{tabular}{||c|c|c|c|c|c||}
\hline
$k_{1}/k_{2}$ & $1$ & $2$ & $3$ & $4$ & $5$\\\hline%
$1$ & $ 2$& $3.1336110\dots$& $4.26792273\dots$& $5.40272433\dots$& $6.53783213\dots$\\\hline%
$2$ & $2.86638898\dots$& $4$
& $5.13343303\dots$& $6.26722203\dots$& $7.40138080\dots$\\\hline%
$3$ & $3.73207726\dots$& $4.86656696\dots$& $6$& $7.13338416\dots$& $8.26697850\dots$
\\\hline%
$4$ & $ 4.59727567\dots$& $5.73277796\dots$& $6.86661583\dots$& $8$&
$9.1333640\dots$\\\hline%
$5$ &$ 5.46216786\dots$& $6.59861918\dots$& $7.73302150\dots$&
 $8.86663592\dots$& $10$\\
\hline\hline
\end{tabular}
\caption{Values of $R_{\pm}$ as a function of $k_1,k_2$.  }
\end{table}
\end{center}

As a consistency check, we see that if we exchange $k_1,k_2$ and add the corresponding entries we always get $2(k_1+k_2)$.

For example we have that for the case $k_{1,2}=1,2$, the combinations $R_+$ and $R_-$ are given by
\begin{eqnarray}
&&\frac{27}{14}+\frac{1}{42} \sqrt[3]{\frac{1}{2} \left(2859138+90720 \sqrt{7446}\right)}-\frac{7893}{7\ 2^{2/3} \sqrt[3]{2859138+90720
   \sqrt{7446}}}\\
   &&  \frac{57}{14}+\frac{1}{42} \sqrt[3]{\frac{1}{2} \left(-2859138+90720 \sqrt{7446}\right)}-\frac{7893}{7\ 2^{2/3} \sqrt[3]{-2859138+90720
   \sqrt{7446}}}
\end{eqnarray}
These are irrational.  Also notice that if we scale $k_1, k_2$ and $R_+$ by the same integer value, then equation \eqref{eq:polyvol} scales homogeneously. This means that the zeroes of $R_+$ for $k_1, k_2$ given can be used to compute the zeroes of $sk_1, sk_2$, and the volume of the new Sasaki-Einstein space goes like $1/s$. Here we see the familiar pattern of the orbifolds emerging whereby rescaling the Chern Simons levels results in a decrease of $1/s$ in the normalized volume of the Sasaki-Einstein manifold.

Notice that via this procedure we are able to recover the complete set of R-charges of gauge invariant operators describing the geometry. This does not automatically give us the R-charge of all the fields in the quiver. As we have already argued only the physical observables are computed.
Notice incidentally that this procedure is completely invariant under Seiberg dualities that leave the chiral ring elements and their relations invariant. Indeed, the constraints on generators are the same and the counting of operators is the same. Thus the dualities that were built by hand pass this test.

Finally, we can ask how to do the same thing for the $A_{n-1}$ quivers that originated our discussion. What is important is to notice that we can expand $R_X=1/2+r$ and
$R_Y=1/2-r$ in terms of a single parameter $r$. The residue at the pole is a function such that $R(r) = R(-r)$, because for every operator we can build a reversed operator where we change the sign of the flux and exchange $X$ fields by $Y$ fields. Thus the sum is symmetric on $r\to -r$.  This function therefore has a critical point at $r=0$. This shows that
we would recover the correct R-charges of the $\mathcal{N}=3$ theory even if we can not count the operators directly.

There is a final remark to be made. This is about how to interpret the volume minimization procedure. In some sense the pole residue counts the
asymptotic growth of the series in $t$, this is, the number of operators with dimension less than some value. Volume minimization tells us to have as few operators as possible below some given dimension. The procedure of volume minimization is then equivalent to trying to maximize the R-charges of the elements of the chiral ring, modulo the fact that there are relations in the chiral ring and some known values of these R-charges as determined from the superpotential equations.

This seems natural from the point of view of the way we think about ordinary field theory where the naive guess is that at strong coupling the anomalous dimensions of operators should get very large.

\begin{center}
\begin{table}
\begin{tabular}{||c|c|c|c|c|c||}
\hline
$k_{1}/k_{2}$ & $1$ & $2$ & $3$ & $4$ & $5$\\\hline%
$1$ & $ 6.84907671\dots$& $4.55476802\dots$& $3.40547919\dots$& $2.71765723\dots$& $2.26040543\dots$\\\hline%
$2$ & $ 4.55476802\dots$& $3.42453835\dots$
& $2.73719476\dots$& $2.27738400\dots$& $1.94888133\dots$\\\hline%
$3$ & $3.40547919\dots$& $2.73719476\dots$& $2.28302556\dots$& $1.95599146\dots$& $1.70988997\dots$
\\\hline%
$4$ & $ 2.71765723\dots$& $2.27738400\dots$& $1.95599146\dots$& $1.71226917\dots$&
$1.52159944\dots$\\\hline%
$5$ &$ 2.26040543\dots$& $1.94888133\dots$& $1.70988996\dots$&
 $1.52159944\dots$& $1.36981534\dots$\\
\hline\hline
\end{tabular}
\caption{Values of normalized volume as a function of $k_1,k_2$. For comparison, the volume of the 7-sphere is
$32.469...$. }\label{tab:vols}
\end{table}
\end{center}

We also have a table of the volumes of the 7-manifolds in \ref{tab:vols}. Notice how all normalized volumes are smaller than the volume of the 7-sphere. Here again, we find that the ratio of volumes of the Sasaki-Einstein seven dimensional spaces to that of the seven sphere is given by an algebraic number, very similar to how it happens in four dimensions \cite{Martelli:2006yb}. The one difference with four dimensions is that in order to solve for the volume we need to find the real root of a cubic equation, rather than roots of quadratic equations. It should also be noted tat most of these are irregular Sasaki Einstein spaces about which very little is known. For five dimensional Sasaki-Einstein spaces many irregular solutions are known, and explicit constructions are given in \cite{Gauntlett:2004yd} ( see also \cite{Martelli:2008rt} for some 7-dimensional Sasaki-Einstein examples).

\section{Conclusion}\label{sec:conclusion}

In this work we study relevant aspects for the computation of moduli spaces of vacua of supersymmetric three dimensional CS theories with matter. On the one hand we provide a method to compute these moduli spaces with semiclassical techniques in a wide class of cases. More specifically, in any superconformal CS-matter theory that has at least $\mathcal{N}=2$ SUSY in three dimensions and quiver matter content ( we consider theories only with bifundamental fields). Our method, based mainly on a generalization of the techniques in \cite{Berenstein:2001jr,Berenstein:2002ge,Berenstein:2007wi}, applies to  theories arising from M2-branes probing various CY four-fold singularities (our methods apply also to non-toric geometries). One of the advantages of our proposed semiclassical analysis is the fact that we do not need the explicit form of the K\"ahler potential in the IR. Also we can compute the spectrum of BPS monopoles operators (for recent work on similar computations, for M2-branes probing $C(Y^{p,q}(\mathbb{CP}^{2}))$ geometries see \cite{Benini:2011cm}). With this information, in principle, by using the volume minimization proposal \cite{MSY}, we can compute exactly all the R-charges in the IR of the operators in the chiral ring. The volume is computed from information of the chiral ring alone, so it does not depend on additional input from gravity or from toric geometry. However this requires a case-by-case analysis and still does not guarantee that one can obtain the exact R-charge for all the fundamental fields (their individual R-charges are not obviously gauge invariant as the fields themselves are not).  That is still an open question for 3d superconformal theories in general. Promising work in that direction is the proposed Z-extremization \cite{Jafferis:2010un}. It would be interesting to compare the results from Z-extremization to what we have obtained here for the $\BC^3/\BZ_3$ quiver. However, there are known problems implementing this proposal when the quiver is chiral \cite{Jafferis:2011zi}.

A novelty of our interpretation of the volume minimization proposal is that it can be described using different language in quantum field theory terms.
The main idea is that the
volume minimization is equivalent to requiring that the number of operators with R-charge less than some value is as small as possible for asymptotically large values of this constant. There are constraints on these dimensions arising from the superpotential, so one can not send all R-charges to infinity. This is, anomalous dimensions should be as large as possible while being compatible with other field theory constraints. This seems very reasonable as it matches a lot of intuition of strong coupling dynamics.

Another aspect we study is the possibility of a Seiberg duality type of transformations that maps one of these three dimensional theories into another. The universality classes are determined by matching holomorphic low energy data: the full moduli spaces of vacua and also the spectrum of dimensions of operators of the chiral ring of the theories. This includes non-perturbative operators. Here, it is worth noticing that if we get the same chiral ring, then the volume minimization procedure will give the same result on both theories.
Since the topology and geometry of moduli spaces depends on the precise values of the Chern-Simons couplings, this is also data that needs to be given in advance before a duality can be claimed. Our approach was to study quiver theories that are related by dualities in four dimensions and to dimensionally reduce and add Chern-Simons terms to the lagrangian. The first part guarantees that the mesonic single trace operators match, but the rest of the analysis of monopole operators depends on details.

A first consistency check for proposed duals of this kind is that their full moduli spaces should match, even in cases where the associated variety is not reduced. In particular, all connected branches to a  given branch should match. For this purpose we first studied a theory with a higher amount of supersymmetry, $\mathcal{N}=3$, so we can have a very detailed control of the dynamics. The theory in question correspond to the reduction of the $A_{n-1}$ theory to three dimensions \cite{Jafferis:2008qz}. In this particular case a brane construction is known \cite{Jafferis:2008qz}. The setup is a generalization of the brane construction in \cite{Kitao:1998mf}. One begins with a D3-brane extended in the $(012)$ directions and wrapped along a compact direction. Call it $x^{6}$. Then $n$ $(1,p_{i})$ 5-branes are put in the $(012[37]_{\theta_{i}}[48]_{\theta_{i}}[59]_{\theta_{i}})$ direction ($i=1,\ldots, n$), intersecting the D3 along $x^{6}$. Then, one way we can arrive at the \emph{dual}, as done in \cite{Amariti:2009rb}, is by moving one $(p,q)$ 5-brane over another, as shown in \cite{Kitao:1998mf}. By the Hanany-Witten effect \cite{Hanany:1996ie}, creation of D3 branes happens as we move one 5-brane over another. This is the set of theories that are presumably dual to each other.
The classical field theory computation we did initially suggested a possible phase transition for  $\mathcal{N}=3$ theories that had non-isolated singularities in the moduli space: the classical computations of the moduli space for the two theories related by a duality did not match. This in turn suggested that the Seiberg-like duality might fail.

One may ask what could go wrong in this picture of the brane construction. When one 5-brane is exactly over another (this is, when they intersect), we have a singularity in the description of the field theory. Suppose we are moving $(1,p_{i})$ over $(1,p_{i+1})$. First the effective Yang-Mills coupling for the $i$-th gauge group goes as $\frac{g_{4}}{L}$, where $g_{4}$ is the 4d coupling and $L$ is the distance between both 5-branes. Then at zero distance we have that $g_{3}$ blows up. This singularity can be avoided for the cases studied in \cite{Hanany:1996ie} by moving the 5-branes in a direction transverse to $x^{6}$ first. The important fact is that this direction must be transverse to both 5-branes, $(1,p_{i})$ and $(1,p_{i+1})$ in our case. For general values of $p_{i}$, $p_{i+1}$ this is clearly not possible, such a direction does not exist. Essentially, what differs is that the singularities in four dimensions happen at single sites in the complex plane and that in three dimensions they happen on the real line. The main assumption about these transformations in four dimensions is that a phase transition can be avoided by taking a contour in the complex plane that avoids the singularity. This can not be done on the real line.
If the potential singularity is unavoidable, we cannot use the trick as for example in \cite{Elitzur:1997fh}, where a brane construction interpretation for Seiberg duality in four dimensional gauge theories is given. Therefore, the theory can in principle undergo a phase transition. Similar observations have also been made in the work \cite{Kapustin:2010mh}.  Also notice that the claim of having a singularity comes from an irrelevant term in the lagranagian (the Yang-Mills lagrangian), so it is possible that this singularity is just an artifact of how the theory is constructed geometrically in terms of branes and not necessarily also a property of the infrared dynamics.

 Happily there are quantum corrections to the quantum numbers of the monopoles that might restore the duality. Indeed, charges of monopole operators suffer quantum corrections (see \cite{Jafferis:2009th,Benini:2009qs} and \cite{Benini:2011cm} for recent work on $\mathcal{N}=2$ theories) and these necessarily change the relations on the chiral ring.  Once these corrections are taken into account, we were able to match the monopole spectrum across the conjectured duals \footnote{We are grateful to C. Closset for showing us his notes detailing this computation}. These corrections depend on the
ranks of the gauge groups and are therefore sensitive to the Hanany-Witten effect, whereas the classical field theory computation does not see these ranks.
A more detailed analysis of Seiberg-like dualities on these cases will be left for future work, as we need a complete understanding of the global properties of the moduli space to establish this fact. Notice also that the existence of fractional fluxes on certain branches \cite{Berenstein:2009sa} also modifies  the naive  structure of the moduli space as a symmetric product and therefore might give a more intricate pattern of dualities when these choices are taken into account.

We have also shown that a change of the CS levels compatible with the action of a reflection functor on the quiver (which can be deduced from topology arguments) in general lead to theories that have totally different moduli spaces, in the case of chiral quivers. This reflection functor technique is a formal way of understanding Seiberg dualities in four dimensional theories that arise form branes in Calabi-Yau geometries. We see that in three dimensions this does not work at all.  To fix it,  we relaxed the analysis, by applying the reflection functor that generates a Seiberg duality to the quiver and matching by hand  the relations between the CS levels of the two theories in order to match the moduli spaces at both sides of the duality. We did this in the particular case of the $\mathbb{C}^{3}/\mathbb{Z}_{3}$ quiver. We found that even if possible dual candidates can be constructed, it is impossible to do it for an arbitrary set of CS levels and some non-trivial congruences and inequalities needed to be satisfied: in general the Chern-Simons levels of the theories can be related by rational coefficients, but not necessarily integer linear combinations. The candidate duals had ranks as prescribed by the reflection functor and without a Hanany-Witten effect.
Needless to say, this is unexpected and deserves further study. In principle we have found new possible Seiberg dual pairs that match at the moduli space level. Because we have non-trivial anomalous dimensions, the dual pairs of  theories are technically strongly coupled and can in principle lead to the same superconformal field theory. On the other hand the recent work \cite{Benini:2011mf}, which came shortly after our paper was originally posted, seems to address some of these issues and has other candidate dual theories that include a Hanany-Witten contribution. A full analysis of the moduli space is still required.

One can assume that one of  the reasons we cannot construct a functor or a duality in a straightforward way, is because we lack of a homological description of M-branes. Therefore one could argue that all  homological operations are suspect. From the point of view of the moduli space computation, we are no longer working in the category of modules of a superpotential algebra (their properties are studied carefully in \cite{Ginzburg:2006fu}). The objects we are working with seem to be  unnatural from the algebraic point of view.  We need $SL$-classes of modules of some quiver algebra instead of the usual $GL$-classes. One possible future direction of this work is to develop a better understanding of these spaces. This may shed some light on a possible extension of the idea of non-commutative resolutions of CY four-folds, analogous to the ones developed in \cite{Berenstein:2000ux,Berenstein:2001jr,Berenstein:2002ge} and  formalized in \cite{VandenBergh} for CY three-folds, but that take into account the new information that M-theory suggests.

\section*{Acknowledgements}

We would like to thank R. Eager, S. Franco and D. Jafferis for many discussions. We are also grateful to C. Closet, S. Cremonesi and A. Kapustin for useful comments after we released the first version of the paper and for alerting us that quantum corrections in the ${\mathcal N}=3$ theories happen.  Work supported in part by DOE under  grant DE-FG02-91ER40618

\appendix

\section*{Appendix}

\section{F-term equations as algebras}

The theories we have considered in this paper have the typical form of quiver theories with superpotential as described from perturbative string theory
techniques. In this appendix we detail the basic considerations that were put forward in the following papers \cite{Berenstein:2000hy,Berenstein:2000ux,Berenstein:2001jr,Berenstein:2002ge,Berenstein:2002fi}. They give a realization of the ideas of \cite{Sharpe:1999qz,Douglas:2000gi} regarding categories for geometric setups in string theory. The main observation is that the typical quiver theories that appear by arguing about perturbative string theory give rise to an associative algebra of relations between the fields. This is a simplified version of thinking about string theory as describing a noncommutative geometry \cite{Witten:1985cc}.

The main setup is as follows. Assume  a  given supersymmetric quiver field theory. The quiver diagram is a directed graph, with the nodes of the quiver representing $U(N_i)$ gauge groups, and the arrows representing chiral superfields in the bifundamental representation $(N_i, \bar N_j)$, which we will collectively call $\phi_{ij}$,  of the two nodes that are connected by the arrow. The arrow direction indicates which of the fields is in the fundamental and which field is in the antifundamental. These can be though of as matrices.

Given the nodes, we define orthogonal projectors $\pi_i$. These objects satisfy the following relations
\begin{eqnarray}
\pi_i \pi_j &=& \delta_{ij} \pi_j\\
\sum_i \pi_i &=& 1\\
\pi_i \phi_{jk} &=& \delta_{ij} \phi_{jk}\\
\phi_{jk} \pi_i &=& \delta _{ik} \phi_{jk}
\end{eqnarray}
The second line indicates that we require the algebra to have an identity.
Using the $\phi_{ij}$ and the $\pi_k$ we can form an algebra of all paths in the quiver. These are formal sums of products of fields and projectors subject to the relations above. It is easy to show that $\phi_{jk} \phi_{\ell m}= \delta_{k\ell}\phi_{jk} \phi_{\ell m}$ by inserting projectors in the middle.
The multiplication is such for two letters joining each other, that a fundamental index is contracted with an antifundamental index. This is just the usual matrix multiplication. Thus, one can think of the fields as arising
from one big matrix where the projectors are giving a block diagonal decomposition, with each block having rank $N_I$, and the arrows $\phi_{ij}$ give matrices
whose only no-zero entries are in the ${ij}$ off-diagonal block.

A superpotential is associated to a perturbative string theory setup if it is generated by a single trace function. The F-term equations that follow from such a function can be shown to give rise to polynomial equations governed by matrix multiplication. These relations are independent of having assigned a rank to each gauge group. Indeed, the main idea is to forget that there is a prescribed rank for each node and to analyze the full setup in terms of the algebra produced this way.

A solution of the F-term equations is then a representation of this algebra. We will label these by $R_\alpha$. Various lemmas follow, which can be found in the papers above.

First, we can take direct sums of representations,  $\oplus R_\alpha$ so given some solutions for some ranks, we can produce solutions for larger ranks by combining information from known solutions. The smallest representations (irreducibles) then serve as building blocks to construct general representations.

If the algebra is regular (this is a homological condition), it is easy to argue that it gives rise to a generalization of a Calabi-Yau three-fold \cite{Berenstein:2002fi}.
The general representation is then essentially a sum of irreducibles. This gives to the full moduli space a structure of a generalized Hilbert scheme of $m$ objects on a (non-commutative) geometry. The D-brane interpretation is that each of these irreducibles is a brane, and the moduli space described many branes on a geometry.

If the algebra is finitely generated over the center, then under suitable conditions the center is a singular Calabi-Yau commutative geometry. The branes in the bulk are
special, and they should only split at singularities of the Calabi-Yau space. This splitting at singularities is described by brane fractionation.
The non-commutative algebra then serves as a resolution of the singularities from a D-brane point of view \cite{Berenstein:2001jr,Berenstein:2002ge}. Many of these results have been formalized in
\cite{VandenBergh}.

The addition of D-terms, or the inclusion to other degrees of freedom generically require to consider a $C^*$ algebra, instead of just a holomorphic object (see \cite{Berenstein:2009ay} for examples of this point of view). Given this extra structure it is always possible to forget the additional generators and relations due to having an adjoint operation. This lets us consider any representation of the bigger object as a representation of the algebra of holomorphic functions. Thus, the full structure of a theory can be considered as a fibration over the F-term algebra.

Given an algebra, one can consider the category of modules of the algebra (these are the representations modulo gauge invariance) and their derived categories. Seiberg dualities in four dimensions can then be described by stating that two such algebras have the same derived category of modules \cite{Berenstein:2002fi}.

For the three dimensional examples in this paper the full moduli space is four-dimensional, but it has a three dimensional Calabi-Yau structure over which it is fibered.
The claim of this paper is that general Seiberg dualities generated by the homological operations that work in four dimensions, generally do not work when considered in three dimensions in the presence of Chern-Simons terms. Given our description above as a fibration over a three dimensional Calabi-Yau geometry, and  that the Calabi-Yau three-fold geometry is invariant, it must be the case that the fibration changes its topology when considering two such theories.


\end{document}